\documentclass[twocolumn,secnumarabic,amssymb, superscriptaddress, nobibnotes, aps, prc]{revtex4-2}
\usepackage{mathtools}
\usepackage[percent]{overpic}
\usepackage{color}
\usepackage{soul,color}
\usepackage{contour}
\usepackage{xurl}
\usepackage[hidelinks]{hyperref}
\hypersetup{
  breaklinks=true, 
  colorlinks   = true,
  urlcolor     = blue,
  linkcolor    = blue,
  citecolor   = blue 
}

\begin{document}

\title{Photo-nuclear cross sections on $^{197}$Au, an update on the gold standard}
\author{J. Song}
\email{songj@anl.gov}
\affiliation{Physics Division, Argonne National Laboratory, Lemont, IL 60439 USA}
\author{D. Rotsch}
\affiliation{Physics Division, Argonne National Laboratory, Lemont, IL 60439 USA}
\affiliation{Radioisotope Science and Technology Division, Oak Ridge National Laboratory, Oak Ridge, TN 37830, USA}
\author{J. A. Nolen}
\affiliation{Physics Division, Argonne National Laboratory, Lemont, IL 60439 USA}
\author{R. Gampa}
\affiliation{Physics Division, Argonne National Laboratory, Lemont, IL 60439 USA}
\author{R. M. de Kruijff}
\affiliation{Physics Division, Argonne National Laboratory, Lemont, IL 60439 USA}
\author{T. Brossard}
\affiliation{Chemical $\&$ Fuel Cycle Technologies Division, Argonne National Laboratory, Lemont, IL 60439 USA}
\author{\\C. R. Howell}
\affiliation{Department of Physics, Duke University, Durham, NC 27708-0308, USA}
\affiliation{Triangle Universities Nuclear Laboratory, Durham, NC 27708-0308, USA}

\author{F. Krishichayan}
\affiliation{Department of Physics, Duke University, Durham, NC 27708-0308, USA}
\affiliation{Triangle Universities Nuclear Laboratory, Durham, NC 27708-0308, USA}

\author{Y. K. Wu}
\affiliation{Department of Physics, Duke University, Durham, NC 27708-0308, USA}
\affiliation{Triangle Universities Nuclear Laboratory, Durham, NC 27708-0308, USA}

\author{S. Mikhailov}
\affiliation{Triangle Universities Nuclear Laboratory, Durham, NC 27708-0308, USA}

\author{M. W. Ahmed}
\affiliation{Triangle Universities Nuclear Laboratory, Durham, NC 27708-0308, USA}
\affiliation{Department of Mathematics and Physics, North Carolina Central University, Durham, NC 27707, USA}

\author{R. V. F. Janssens}
\affiliation{Triangle Universities Nuclear Laboratory, Durham, NC 27708-0308, USA}
\affiliation{Department of Physics $\&$ Astronomy, University of North Carolina at Chapel Hill, Chapel Hill, NC 27599-3255, USA}

\date{\today}

\begin{abstract}
  Cross sections for the $^{197}$Au$(\gamma,n)$ reaction are broadly used in nuclear
  physics as a standard for normalizing photonuclear reaction cross-section data at photon beam
  energies above approximately 8 MeV.
  In this paper, we report cross-section measurements for the
  $^{197}$Au$(\gamma,n)^{196}$Au$^{g+m1}$ reaction at beam energies from 13 to 31 MeV.
  Our measurements provide the first cross-section data for this reaction at beam energies above 20 MeV,
  enabling the use of this reaction as a cross-section standard up to 30 MeV.
  Also, this work provides first cross-section measurements for the $^{197}$Au$(\gamma,n)^{196}$Au$^{m2}$ reaction.
  In addition, we measured cross-section data for the $^{197}$Au$(\gamma,3n)^{194}$Au reaction,
  which can be used as a cross-section standard above about 25 MeV.
  These measurements were performed using a new target activation method that is based on the
  angle-energy correlation of the laser Compton-scattered photon beams at the High Intensity
  Gamma-ray Source (HI$\gamma$S).
  The technique enables measuring photonuclear reaction cross-sections
  at several discrete beam energies concurrently via a single irradiation on a stack of different targets.
  Measurements were carried out by irradiating a stack of concentric-ring targets
  consisting of Au, TiO$_2$, Zn, Os, and Au (in order of the $\gamma$-ray beam direction).
  Our data for the $^{197}$Au($\gamma$,n)$^{196}$Au$^{g+m1}$ reaction in the energy range of 13 to 20 MeV are
  in good agreement
  with existing ones measured using monoenergetic $\gamma$-ray beams, but differ from data acquired
  using a bremsstrahlung
  $\gamma$-ray beam.
  Also, above 18 MeV, our data for the $^{197}$Au($\gamma$,n)$^{196}$Au$^{g+m1}$
  and $^{197}$Au($\gamma$,n)$^{196}$Au$^{m2}$ reactions differ significantly
  from the most recent TENDL and JENDL evaluations, suggesting a need to update these data libraries.
  The TENDL evaluation and existing data are consistent with our data for the $^{197}$Au($\gamma$,3n) reaction,
  but differ significantly from the JENDL evaluation above 26 MeV.

\end{abstract}

\maketitle

\section{Introduction}
\label{introd:int}
Radioisotopes play important roles in numerous fields ranging from medical treatments to national security and basic research. Some examples include
investigations of structures and reactions involving atomic nuclei, Mossbauer spectroscopy, radio-thermoelectric generation and other nuclear batteries,
nuclear device detection, and the mitigation of nuclear proliferation \cite{int1}.  The reports from “Workshop on the Nation’s Needs for Isotopes: Present and Future”,
“Isotopes for the Nation’s Future, A long range plan”, and “Meeting Isotope Needs and Capturing Opportunities for the Future” identify multiple isotopes
currently in short supply \citep{int1,int2,int3}. Among these are high specific activity beta emitters such as $^{47}$Sc, $^{67}$Cu, $^{77}$As, and $^{186}$Re.
These radioisotopes are of interest
to the community as they have ideal nuclear properties for medical applications. Their availability is limited as they are difficult to produce with high
specific activity using common production methods.

Most radioisotopes are produced using either nuclear reactors or light-ion accelerators. 
High specific activity radioisotopes are very important in nearly all aspects of radiochemical work.
Specific activity refers to the radioactivity of the produced radionuclide relative to the total mass of the element of
interest present in the sample.  
Most radionuclides produced in nuclear reactors have low specific activity due to the inherently
limited production pathways, e.g., dominated by (n,$\gamma$) reactions.
In contrast, radionuclides produced at particle accelerator facilities via charged-particle induced reactions
are generally characterized by high specific activity, e.g., (p,xn)
reactions are typically used because the produced isotope is a different element from the target material,
allowing chemistry techniques for separating materials.  Also, photonuclear reactions, e.g., ($\gamma$,x),
provide a promising process for producing high specific activity radioisotopes.  
The Nuclear Science Advisory Committee (NSAC) recently identified the use of photonuclear reactions
as one of the most compelling opportunities for large-scale  production of high specific activity radioisotopes
using high-flux bremsstrahlung $\gamma$-ray beams generated with high-current linear accelerators (LINACs) \cite{int1}. 
The isotope production rate is directly proportional to the integral of the product of the $\gamma$-ray beam flux on the target
and the photonuclear cross section. The end-point energy of the bremsstrahlung $\gamma$-ray beam spectrum is usually adjusted
to optimize isotope production over the Giant Dipole Resonance (GDR) energy region.   Accurate cross-section data
for photonuclear reactions over this GDR energy region are needed for developing strategies for radioisotope production
with high specific activity.
A great deal of work has been done on measuring experimental cross-section databases for some reactions,
e.g., ($\gamma$,n) reactions \citep{int4,int5,int6}.  However, more measurements are needed.
For example, the cross-section data for ($\gamma$,p) reactions (which lead to high specific activity radioisotopes)
still require considerable work. There are many theoretical databases available which provide guidance
on photo-production of radionuclides. Experimental data are needed to evaluate the accuracy of predictions
of the leading reaction models used in nuclear-data libraries (TENDL, JENDL, ENDF, JEFF, CENDL, BROND, etc.)
\citep{int7,int8,int9,int10,int11,int12,int13,int14,int15,int16}. 
Examples of theoretically predicted cross sections which incorrectly estimate the production rate of radioisotopes 
are reported in the literature \citep{int17,int18,int19,int20}.
Photonuclear cross sections are typically measured using broad energy spectrum bremsstrahlung $\gamma$-ray beams
(either untagged or tagged) or quasi-monoenergetic $\gamma$-ray beams produced via laser Compton backscattering (LCB)
or positron in-flight annihilation.
For quasi-monoenergetic beams, LCB is the modern beam production method, see Table \ref{tab:Au_data}.
Compared with a bremsstrahlung beam, which has a broadband energy intensity distribution, a LCB $\gamma$-ray beam
is narrowly peaked around a specific energy. Precise measurement of cross sections can be performed
using activation techniques with target irradiation by such LCB $\gamma$-ray beams.  
Independent of the type of $\gamma$-ray beam used, cross-section measurements require accurate determination of the beam flux.
One approach is to establish a cross-section standard that can be used to determine the beam flux over
a specified beam energy range.
The cross sections for the $^{197}$Au($\gamma$,xn) reactions are widely used as a standard for determining the photon beam flux
in photonuclear reaction measurements. As a standard for cross-section normalization, errors in $^{197}$Au($\gamma$,xn)
cross-section data will impact a broad range of nuclear physics research and applications.
For example, gold monitor foils are routinely used to normalize bremsstrahlung photon beams in cross-section measurements
\citep{int25,int26,int27}.
The current status of $^{197}$Au($\gamma$,x) cross-section data is as follows.
The $^{197}$Au($\gamma$,n)$^{196}$Au cross section data from different experiments are in good agreement in the energy range of 8-14 MeV,
which is below the threshold for multiple neutron emission (see Table \ref{tab:Au_data}).
However, there are significant discrepancies between data \cite{1,2} at photon energies between 14-18 MeV.
Also, there are no cross-section data for the $^{197}$Au($\gamma$,n) reaction above 20 MeV
and extrapolation of data evaluations by TENDL and JENDL differ substantially.  

In this paper, we describe our  newly developed experimental techniques for measuring photonuclear cross sections
at discrete energies using a LCB $\gamma$-ray beam and its application to measuring 
cross sections for photon-induced reactions on $^{197}$Au.
The photon energies in this work were chosen to be above the threshold for reactions leading to high specific activity radionuclides via
the ($\gamma$,p) reaction for use in medical applications such as $^{48}$Ti($\gamma$,p)$^{47}$Sc and $^{68}$Zn($\gamma$,p)$^{67}$Cu
at around 20 MeV and higher.
The technique is based on the well-defined radial energy dependence of a LCB $\gamma$-ray beam which enables measurement
of the corresponding spatially dependent cross section.
The central portion of the $\gamma$-ray beam has the highest energy
and flux density, with the beam energy and flux density decreasing radially outward from the center in a predictable manner \cite{beam1}.
Activation of concentric ring targets by an uncollimated beam will provide experimental activation data to be used
to determine cross sections at the average energy of the beam incident on each ring. This method was developed
and used for the first time at the Triangle Universities Nuclear Laboratory’s (TUNL) High-Intensity Gamma-Ray Source
(HI$\gamma$S). The HI$\gamma$S facility is described by Weller et al. in Ref.  \cite{int23}. 
A goal of the present work is to measure the exclusive cross-section excitation functions of the $^{197}$Au($\gamma$,n)
and ($\gamma$,3n) reactions for use as cross-section standards.
Table \ref{tab:Au_data} provides a summary of the world data for the $^{197}$Au($\gamma$,xn) reactions by beam type.
An interesting feature of the existing data is that there is apparently a strong correlation between the beam type
(quasi-monoenergetic or bremsstrahlung) and the experimental technique. All experiments in Table \ref{tab:Au_data}
using a quasi-monoenergetic $\gamma$-ray beam measured the cross section with direct neutron counting.
The measurements using a bremsstrahlung $\gamma$-ray beam determined the cross section via activation techniques.
Our measurements are the first to be done with a monoenergetic $\gamma$-ray beam while using activation techniques.
In Section \ref{xs:result}, our data are compared to the results of the previous measurements listed in Table \ref{tab:Au_data}.
The results of this work for the Ti, Zn, and Os targets measured concurrently with Au will be presented in a subsequent paper.
Here, our multiple target  multi-energy technique is described in section \ref{method:int}, the HI$\gamma$S LCB
 beam is presented in section \ref{qm:gamma},
the experiment is discussed in section \ref{expt:tgt}, section \ref{xs:result} describes the cross-section analysis, and
the results are presented and discussed in section \ref{RD:result}.
\begin{table*}
  \caption{\label{tab:Au_data} World data for the $^{197}$Au($\gamma$,xn) cross section measurements by beam type.}
  \footnotesize
  \begin{ruledtabular}
    \begin{tabular}{ cr rr rr rr}
       Reaction & Reference & Year &$E_{\gamma}$ range (MeV) & Points & Lab & Beam type & Technique\\
       \hline
       ($\gamma$,n)	&	C. Plaisir et al. \cite{1}	&	2012	&	10-20	&	101	&	CEA/DAM ELSA, France	&	Bremsstrahlung	&	activation counting	\\
($\gamma$,n)	&	K. Vogt et al. \cite{9}	&	2002	&	8.08-9.90	&	9	&	TU-Darmstadt, Germany	&	Bremsstrahlung	&	activation counting	\\
($\gamma$,n)	&	B.L. Berman  et al. \cite{8}	&	1987	&	12.1-16.9	&	11	&	LLNL, USA	&	$e^{+}$ annihilation	& neutron counting	\\
($\gamma$,n)	&	A. Veyssiere et al. \cite{7}	&	1970	&	8.08-19.8	&	43	&	CEA Paris-Saclay, France	&	$e^{+}$ annihilation	&	neutron counting	\\
($\gamma$,n)	&	S.C. Fultz et al.  \cite{2}	&	1962	&	8.7-17.9	&	30	&	AIST, Tsukuba, Japan	&	$e^{+}$ annihilation	&	neutron counting	\\
      ($\gamma$,n)	&	O. Itoh et al. \cite{5}	&	2011	&	8.08-13.14	&	30	&	AIST, Tsukuba, Japan	&	LCB source	&	neutron counting	\\
      ($\gamma$,n)	&	F. Kitatani et al. \cite{6}	&	2011	&	9.3-12.4	&	12	&	AIST, Tsukuba, Japan	&	LCB source	&	neutron counting	\\
($\gamma$,n)	&	F. Kitatani et al. \cite{4}	&	2010	&	10.0-14.6	&	8	&	AIST, Tsukuba, Japan	&	LCB source	&	neutron counting	\\
      ($\gamma$,n)	&	K.Y. Hara  et al. \cite{3}	&	2007	&	8.23-12.3	&	24	&	AIST, Tsukuba, Japan	&	LCB source	&	neutron counting	\\
      \hline
      ($\gamma$,2n)	&	B.L. Berman  et al. \cite{8} 	&	1987	&	14.0-16.9	&	7	&	LLNL, USA	& $e^{+}$ annihilation	&	neutron counting	\\
($\gamma$,2n)	&	A. Veyssiere et al. \cite{7}	&	1970	&	13.5-27.1	&	40	&	CEA Paris-Saclay, France	&	$e^{+}$ annihilation	&	neutron counting	\\
      ($\gamma$,2n)	&	S.C. Fultz et al. \cite{2}	&	1962	&	14.3-24.7	&	29	&	LLNL, USA	&	$e^{+}$ annihilation	&	neutron counting	\\
      \hline
      ($\gamma$,3n)	&	A. Veyssiere et al. \cite{7}	&	1970	&	22.8-27.1	&	8	&	CEA Paris-Saclay, France	&	$e^{+}$ annihilation	&	neutron counting	\\
      
    \end{tabular}
  \end{ruledtabular}
\end{table*}

\section{Method}
\label{method:int}
This section provides details about the method applied in this work to measure the exclusive
($\gamma$,n) and ($\gamma$,3n) cross sections via the activation method.
The excitation functions reported here cover the $\gamma$-energy range from 13 to 31 MeV.
The photonuclear cross sections over the range of the Giant Dipole Resonance (GDR) vary by orders of magnitude.
The goal of this work is to demonstrate our activation techniques by applying it in measuring the excitation functions
of the $^{197}$Au($\gamma$,xn) reaction cross sections.
with an energy resolution of 3-10$\%$ at each energy.
At HI$\gamma$S, the beam energy spread depends on the diameter of the beam defining collimation.
In a typical setup with an energy spread (FWHM) of 5$\%$ (FWHM),
the flux transmitted through the collimator is about 7.5$\%$ of the total flux from the Compton-scattered laser beam.
Our method of using three radial target segments enables
concurrent measurement of the cross section at three central beam energies.
The radial energy and flux distribution of the HI$\gamma$S beam was verified experimentally by blocking portions of
the beam with tungsten plugs to mimic the concentric-rings target design.
Simulations were performed and compared to the experimental results of these calibrations to determine the accuracy of the beam simulations.

\subsection{Target stack}
\begin{figure}[h!]
  \begin{center}
    \begin{overpic}[width=0.45\textwidth]{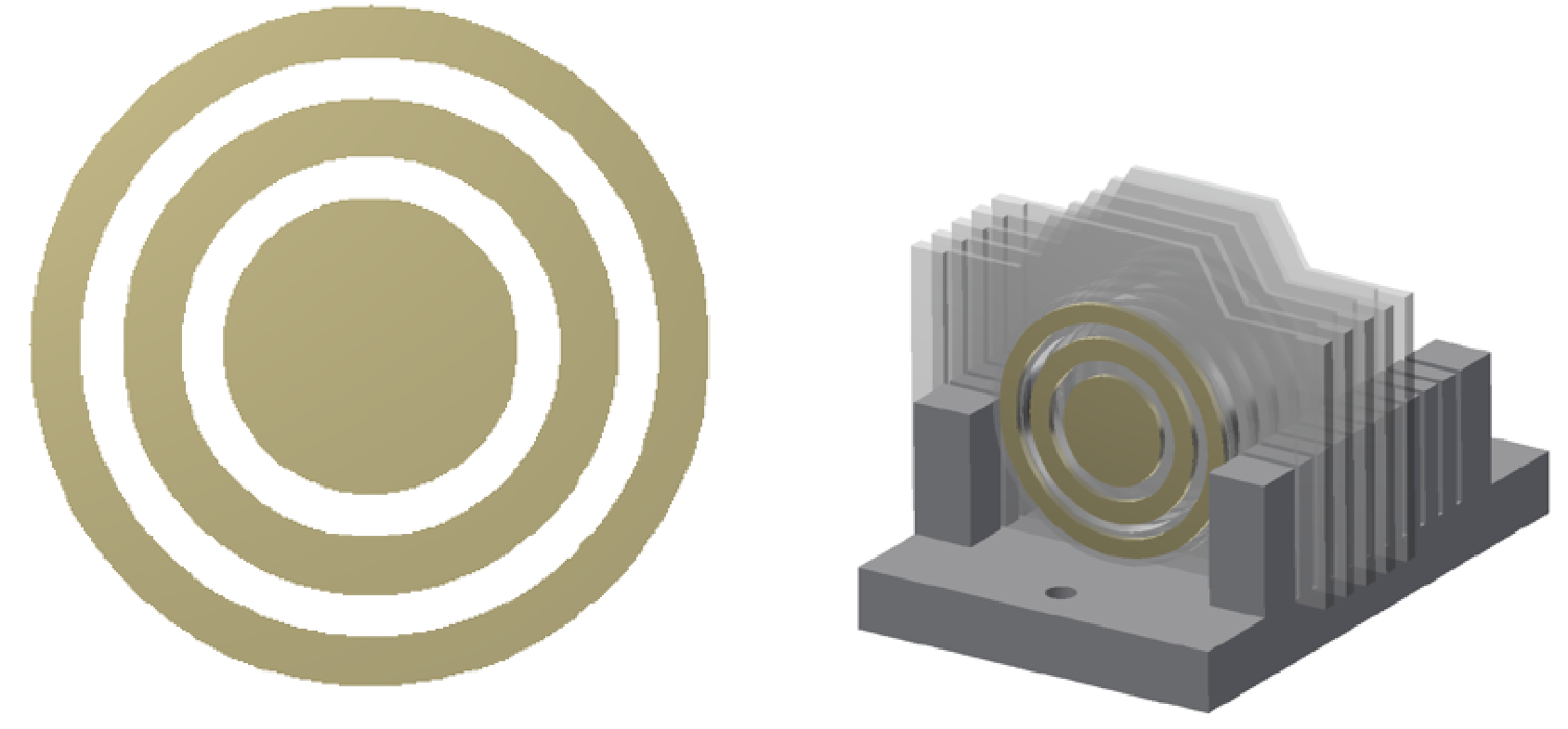}
      \put(24,25){\vector(1,0){8}}
      \put(24,25){\vector(1,1){10}}
      \put(24,25){\vector(0,1){20}} 
      \put(26,44){\scriptsize $3^{rd}$ seg}
      \put(34,31){\scriptsize $2^{nd}$ seg}
      \put(22,21){\scriptsize $1^{st}$ seg}
      \put(62,37){\makebox(25,5){\downbracefill}}
      \put(62,42){\scriptsize multiple targets}
    \end{overpic}
    \caption{Schematic design of the target stack. A stack of five targets, three-fold segmented, was used for the experiments.
      The inner(outer) diameters of the three segments in each target are 0(1.42), 1.82(2.38) and 2.78(3.26) cm, respectively.}
    \label{fig:tgt}
  \end{center}
\end{figure}
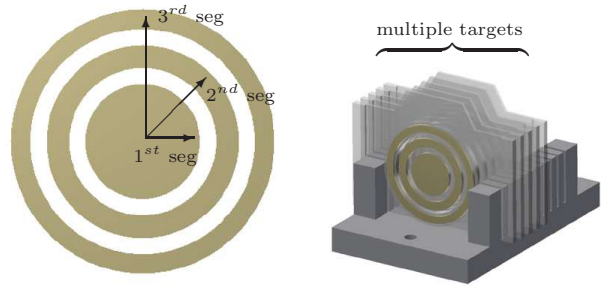

In the activation method, the photon beam induces radioactivity via photon-induced reactions, such as ($\gamma$,n) and ($\gamma$,p), in targets.
Reaction yields are determined by gamma counting the targets after activation.
The target foils used in these experiments are made of separable concentric rings that exploit the energy spread of the HI$\gamma$S beam
and enable the activation of multiple targets across multiple energies in a single beam exposure.
A concept of a target stack is illustrated in Figure \ref{fig:tgt}. 
The relatively low and well-characterized attenuation of high-energy photons by the target materials provides the ability to irradiate multiple targets in a single irradiation.
The total thickness of the stack of targets used in these experiments was 2.8-3.3 g/cm$^{2}$, which resulted in a 13$\%$
total attenuation of the photon beam at 14 MeV. The entire target may then be exposed to the beam and each individual
concentric ring will interact with a narrow range of beam energies and a calculable fraction of the overall photon flux.
The individual rings are analyzed with high-purity germanium detectors (HPGe) to determine activation yields.
Thus, using this technique, the cross sections at three energies can be measured in a single irradiation.
With the stack of five targets and three segments per target, yields for 15 target segments are obtained for each $\approx$ 12-hour run.

\subsection{Activation}

Using the activation method (see details in Appendix \ref{appdx:act}), the cross sections on the $^{197}$Au targets were obtained.
The combined cross section for ground and $1^{st}$ isomeric state is given by,
\begin{equation}
  \sigma_{g+m1} = \dfrac{\lambda_{g}(N_{g}^{evt}- \beta -  \gamma)}
        { N_{t} I (1-e^{-\lambda_{g} t_{0}})e^{-\lambda_{g} t_{c}}(1-e^{-\lambda_{g}t_{m}})}
        \label{eq:5}
\end{equation}
with,
\begin{equation*}
  \footnotesize
  \begin{split}
    \alpha & = \frac{R_{m2}}{\lambda_{g}}(1-e^{-\lambda_{g}t_{0}})-\frac{R_{m2}}{\lambda_{m2} - \lambda_{g}}(e^{-\lambda_{g}t_0{}}-e^{-\lambda_{m2}t_{0}})\\
    \beta & = N_{m2}^{0} \frac{\lambda_{m2}\lambda_{g}}{\lambda_{m2} - \lambda_{g}}
    (\frac{e^{-\lambda_{g}t_{c}}}{\lambda_{g}}(1-e^{-\lambda_{g} t_{m}})-\frac{e^{-\lambda_{m2}t_{c}}}{\lambda_{m2}}(1-e^{-\                                       
\lambda_{m2} t_{m}}))  \\
    \gamma & = \alpha e^{-\lambda_{g}t_{c}}(1-e^{-\lambda_{g} t_{m}})
  \end{split}
\end{equation*}
The cross section for producing $^{196}$Au in the 2nd isomer state is,
\begin{equation}
   \sigma_{m2}  = \frac{\lambda_{m2} N_{m2}^{evt}}{ N_{t} I(1-e^{-\lambda_{m2}t_{0}})e^{-\lambda_{m2} t_{c}}
      (1-e^{-\lambda_{m2} t_{m}})}
  \label{eq1:4}
\end{equation}
The cross section for producing $^{194}$Au is determined in the same way as Equation \ref{eq1:4},
\begin{equation}
  \small
  \begin{split}
    \sigma  = \frac{\lambda N^{obs}}{ N_{t} I(1-e^{-\lambda t_{0}})e^{-\lambda t_{c}}
      (1-e^{-\lambda t_{m}})   }
  \end{split}
  \label{eq:61}
\end{equation}

\section{quasi-monoenergetic Photon beam}
\label{qm:gamma}

\begin{figure}
  \begin{center}
    \begin{overpic}[width=0.49\textwidth]{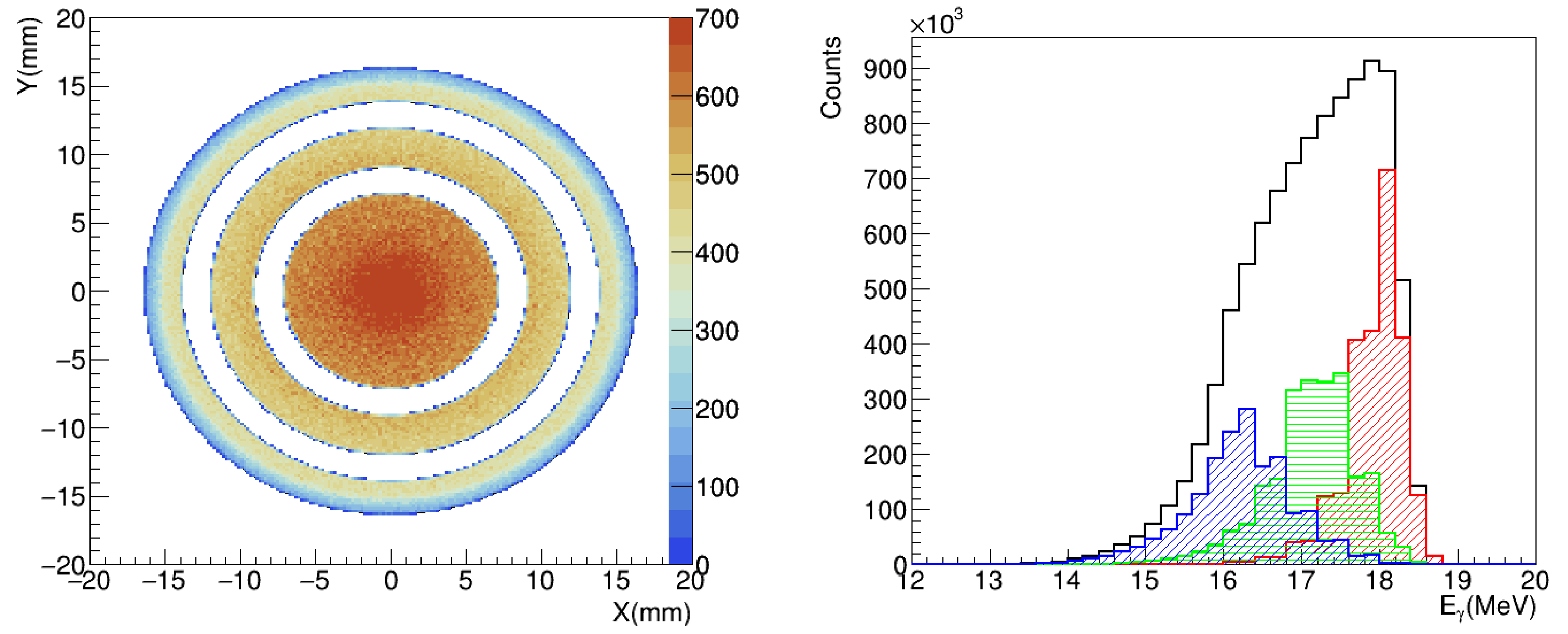}
      \color{red}
      \put(27,22){\vector(1,0){62}}
      \put(60,23){\tiny $1^{st}$, 18 MeV}
      \color{green}
      \put(32,16){\vector(1,0){52}}
      \put(60,17){\tiny $2^{nd}$, 17 MeV}
      \color{blue}
      \put(34,10){\vector(1,0){45}}
      \put(60,11){\tiny $3^{rd}$, 16 MeV}
      \color{black}
      \put(8,34){$a$)}
      \put(61,34){$b$)}
    \end{overpic}
    \caption{Simulated beam energy and flux for the 18-MeV run,
      as an example. a) Spatial distribution of the incident photon beam on the front of the segmented target.
      The inner and outer diameters of the 1$^{st}$, 2$^{nd}$ and 3$^{rd}$ segments were 0 - 1.42 cm,
      1.82 - 2.38 cm and 2.78 - 3.26 cm, respectively.
      b) Photon flux vs energy in each target segment.}
      \label{fig:be}
  \end{center}
\end{figure}
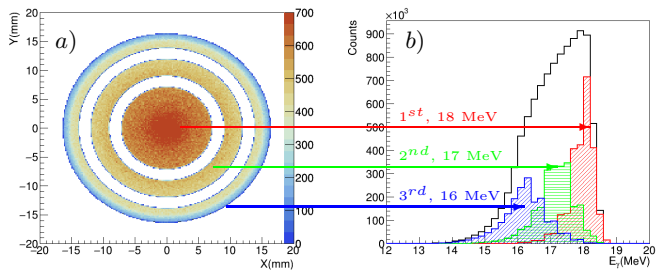

The photon beams are produced by Compton scattering of photons from relativistic electrons
circulating in a storage ring at HI$\gamma$S. In the case of relativistic electrons ($\gamma \gg$ 1)
scattering at a small angle ($\theta_{f}\ll$1), the scattered photon energy after collision is given by,
\begin{equation}
  E_{g} \approx \dfrac{4\gamma^{2}E_{p}}{1+\gamma^{2} \theta_{f}^{2} + 4\gamma^{2}E_{p}/E_{e}} 
\end{equation}
where E$_{g}$ is the scattered photon energy, and E$_{p}$ and E$_{e}$ are the energies of the incident photon and electron.  
The Lorentz factor of the electron is denoted as $\gamma$. The angle $\theta_{f}$ is expressed in terms of the radius of the target (r) and the distance
from the collision point to the target position, r/L. 

\begin{figure}
  \begin{center}
    \begin{overpic}[width=0.168\textwidth]{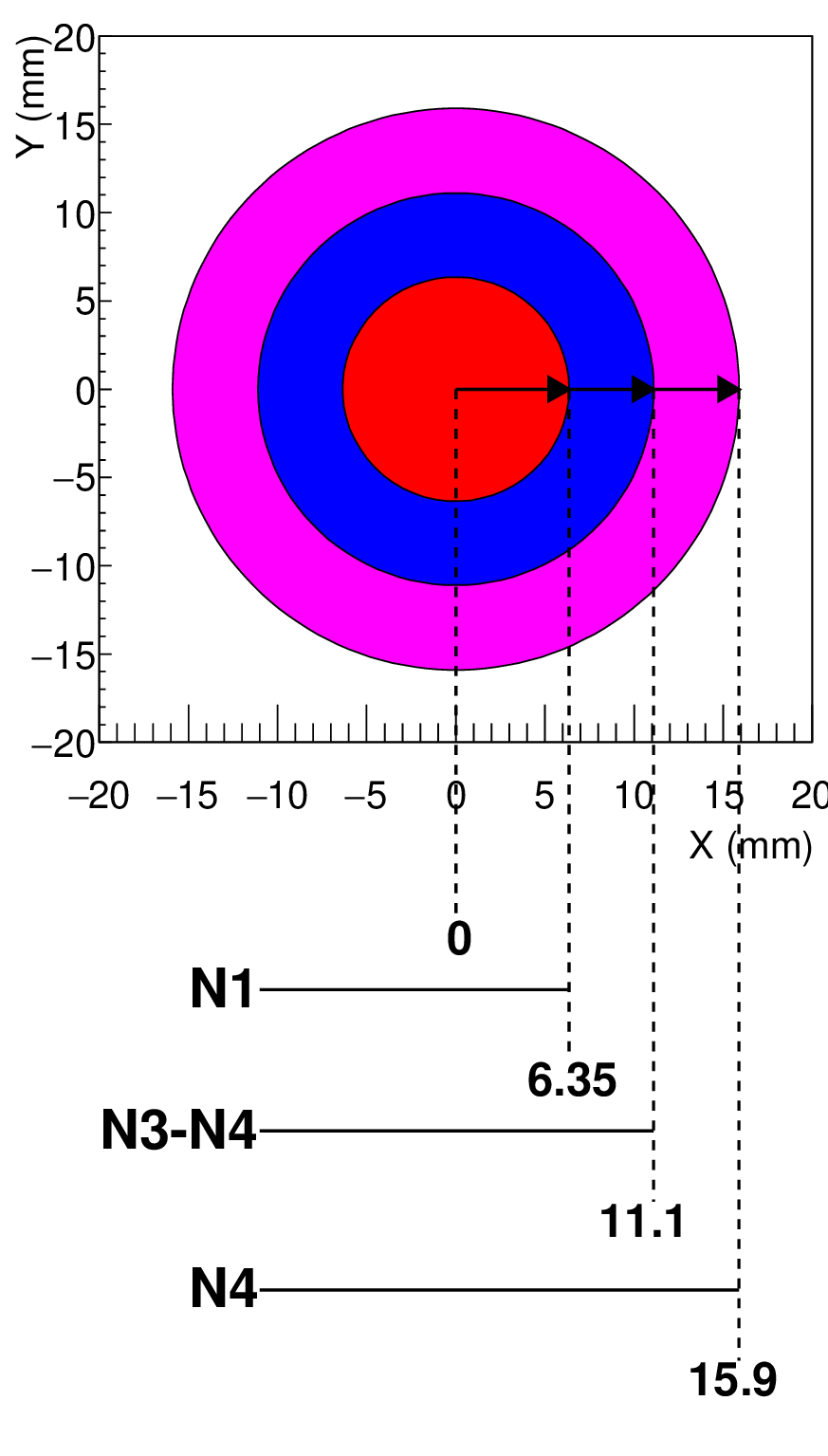}
    \end{overpic}
     \begin{overpic}[width=0.305\textwidth]{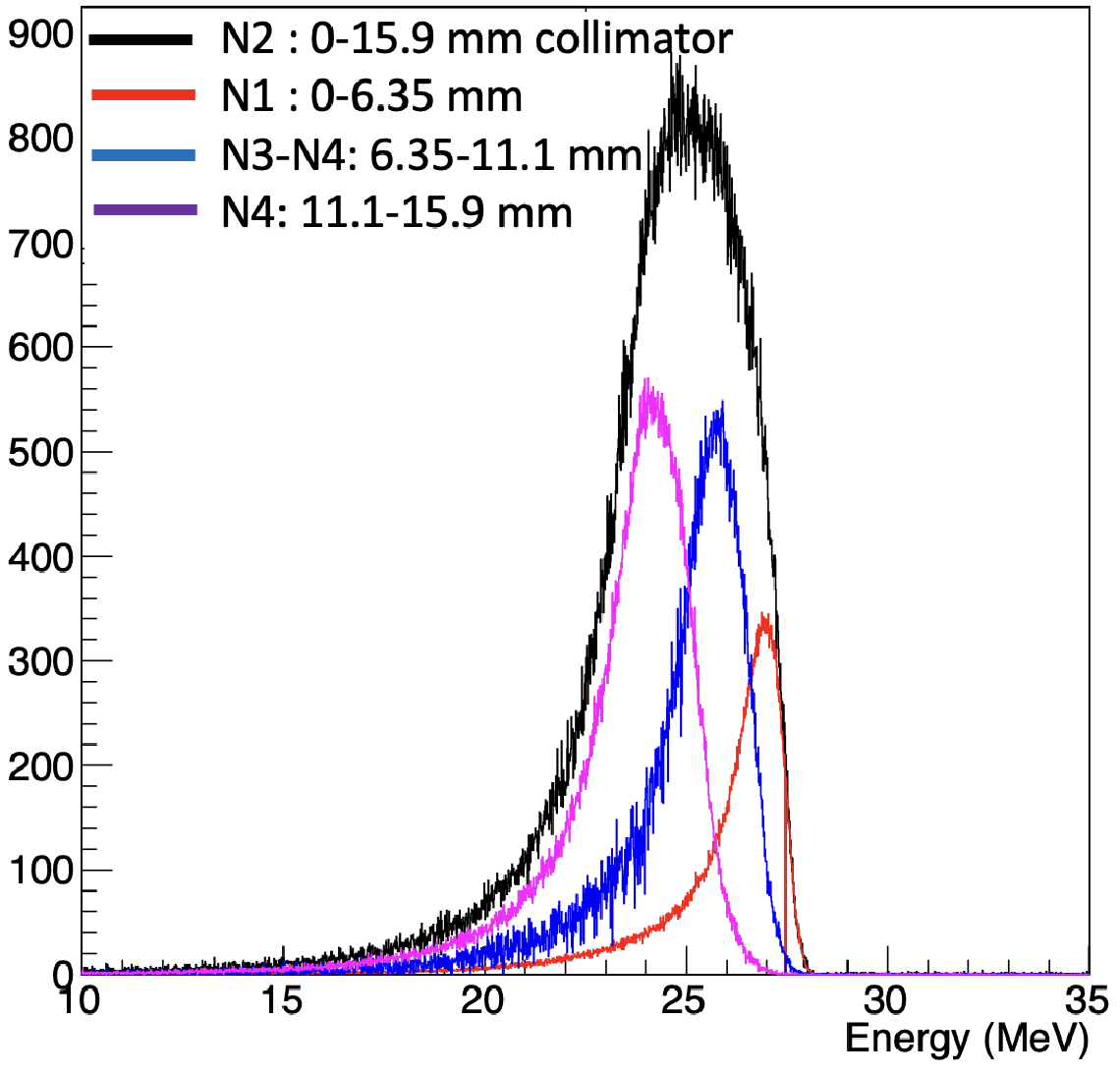}
    \end{overpic}
   \caption{Measurement of I(E$_{\gamma}$) for different collimator
      configurations using Molly. The energy spectra are not corrected for the detector energy resolution,
      which is about 4 $\%$ FWHM (see text for details).} 
   \label{fig:mm}
  \end{center}
\end{figure}

As stated above, for each beam irradiation, reaction cross sections were measured for each target at three average beam energies,
one for each target segment. The cross sections were determined from the measured activation of each target segment using
Equations \ref{eq:4} – \ref{eq:61} (see Appendix \ref{appdx:act}).
The parameter $I_{0i}$ in the equations is the $\gamma$-ray beam flux incident on each target segment,
i.e., $I_{0i} = I_{0} \cdot f_{i}$,
where $I_{0}$ is the total beam flux in the direction of the target and $f_{i}$ is the fraction of $I_{0i}$ incident
on ring i. The uncertainty in $I_{0i}$, $\Delta I_{0i}$,
is the dominant systematic uncertainty in the cross-section measurements.
There are two systematic uncertainties contributing to $\Delta I_{0i}$: (1) the uncertainty in the measurement of $I_{0}$,
and (2) the uncertainty in the calculated value of $f_{i}$.  The $I_{0}$ measurements are described in section \ref{expt:pbf}.
The determination of $f_{i}$ for each segment and its uncertainty is described in this section.
We used the model of Sun and Wu \cite{beam1} to simulate the Compton-scattered photons passing through
the free-electron laser (FEL) aperture systems to obtain the beam intensity profile (i.e., $I_{0}(r)$,
the beam intensity as a function of the radial distance from the beam axis) and the photon energy spectrum
(i.e., $I_{0}(E_{\gamma})$,
the beam intensity as a function of the photon energy $E_{\gamma}$. The calculated $I_{0}(r$) and $I_{0}(E_{\gamma})$
from the simulations have been benchmarked with other codes such as CAIN by Yokoya \cite{beam2}
and using experimental measurements at HI$\gamma$S \cite{beam1}.
An example of the Monte-Carlo simulations for our experiment is given in Figure \ref{fig:be}.
For this simulation, the $\gamma$-ray beam at HI$\gamma$S is produced with $E_{e}$ $\approx$ 660 MeV
and a free electron laser photon wave length $\lambda \approx 450$ nm.
The radial extent of each ring is given in Figure \ref{fig:be}. The energies of peak intensity of the scattered $\gamma$-ray
beam at the target location (L $\approx$ 53 m) incident on the three segments are $E_{\gamma}\approx$ 18, 17 and 16 MeV, respectively.
The beam intensity distributions incident on each ring, $I_{0i}(r)$ and $I_{0i}(E_{\gamma})$,
are plotted in Figure \ref{fig:be} a) and b), respectively.
For the beam energy range of 13 to 31 MeV covered in our measurements,
the typical percentile energy spreads (1$\sigma$) of the photon beam incident on the $1^{st}$, $2^{nd}$ and $3^{rd}$ target rings were
1.3-2.7$\%$, 2.2-4.3$\%$ and 3.2-5.7$\%$ of the central beam energy, respectively.
The $I_{0i}(r)$ and $I_{oi}(E_{\gamma})$ distributions were computed from the simulations
and used to calculate the values of $f_{i}$
and $E_{i}$, the photon energy of the peak intensity of the $I_{0i}(E_{\gamma})$ distribution for each ring.
We performed benchmark measurements to assess the systematic uncertainties in the calculations of $f_{i}$ and $E_{i}$ using the simulated data.   
These measurements were performed using $\gamma$ rays produced under conditions identical to those used in the simulations,
$E_{e} \approx 725$ MeV and $\lambda \approx 350$ nm. Special beam collimator arrangements were used to select sections of the beam
cross section like the rings in the targets.
Each collimator configuration consisted of a lead collimator with an open circular aperture,
or the same lead collimator
with a circular tungsten rod insert.  The lead collimator and tungsten rods were 6 inches (15.2 cm) in length.
The axes of the collimator aperture and the cylindrical rod were centered on the beam axis.
At the central beam energy of about 25 MeV, the $\gamma$-ray attenuation through the body of the lead collimators
and tungsten rods were about $2 \times 10^{-5}$ and $3 \times 10^{-8}$, respectively. 
The energy spectrum of the beam was measured using a large cylindrical NaI detector (10” dia. $\times$ 14” thick.),
referred to as “Molly”.  The efficiency of Molly is greater than 98$\%$ at the energies of these measurements.
For the beam energy profile measurements, Molly was positioned with its axis centered on the $\gamma$-ray beam axis,
and copper attenuators were inserted in the beam path to reduce the beam flux 
incident on Molly to less than about $2 \times 10^{3} \gamma/s$, where the pulse-pileup rate is negligible.
Measurements of $I_{0i}(E_{\gamma})$ were performed for the first four collimator configurations
itemized in Table \ref{fig:be} a): N1 = 0.0 to 6.35 mm; N2 = 0.0 to 15.9 mm; N3 = 6.35 to 15.9 mm; and N4 = 11.1 to 15.9 mm.
These measurements were carried out using two collimators with aperture diameters of 0.5” (12.7 mm) and 1.25” (31.8 mm)
and two tungsten rods with diameters 0.5” (12.7 mm) and 0.875” (22.2 mm).
In addition, the difference in the measured $I(r)$ for two collimator configurations
was determined by comparing measured data to simulations for radial ranges similar to the target rings,
e.g., see configuration 5 in Table \ref{tab:bm}.
The values for $f_{i}$ obtained from the simulated $I_{0i}(E_{\gamma})$ distributions differ from those determined
using the measured $I_{0i}(E_{\gamma})$ distributions by about 2$\%$, see the results
for collimator configurations 1–4 in Table \ref{tab:bm}.  This uncertainty is significantly smaller than the error
in our determination of the absolute beam flux, see section \ref{expt:pbf}.
As shown in Table \ref{tab:bm}, the mean beam energy computed using the simulated data is in good agreement
with the measured mean beam energy in each segment.
Examples of measured $I_{0i}(E_{\gamma})$ distributions for several collimator configurations can be found in Figure \ref{fig:mm}. 

\begin{table*}
  \caption{\label{tab:bm} Summary of the beam collimator configurations used in the benchmark measurements of $I_{0i}(r)$
    described in section \ref{qm:gamma} and the results for $f_{i}$ and $E_{i}$ determined from the measured $I_{0i}(r)$
    compared to the values calculated from the simulated $I_{0i}(r)$.  The collimators are positioned in the standard location
    in the collimator hut at about 50 m downstream from the electron-photon collision point in the storage ring.
    Note: (a) full flux measurement and simulation; (b) computed using configuration 3 and 4 measurements.}
  \begin{ruledtabular}
    \begin{tabular}{r | r  r r r |rrr r|r}
      Coll. & Coll. Dia. & Rod Dia. & $R_1{}$ & $R_{2}$ & Meas. & Sim. & Meas. $E_{i}$ & Sim. $E_{i}$ & Note \\
      Config. & mm & mm & mm & mm & $f_{i}(\%)$ &  $f_{i}(\%)$ & MeV & MeV\\
      \hline
      \hline
      1 &12.7&  none        & 0.0       &6.35   &18     &20     &27.1   &27.1\\
      2 &31.8&  none        & 0.0       &15.9   &100    &100    &25.2   &25.6 & (a)\\
      3 &31.8&  12.7        & 6.35      &15.9   &81     &80     &24.9   &25.6\\
      4 &31.8&  22.2        &11.1       &15.9   &45     &45     &24.1   &24.5\\
      5 &31.8&  12.7/22.2   &6.35       &11.1   &37     &35     &25.8   &26.1&  (b)\\
    \end{tabular}
  \end{ruledtabular}
\end{table*}

\begin{figure}
  \begin{center}
    \begin{overpic}[width=0.30\textwidth]{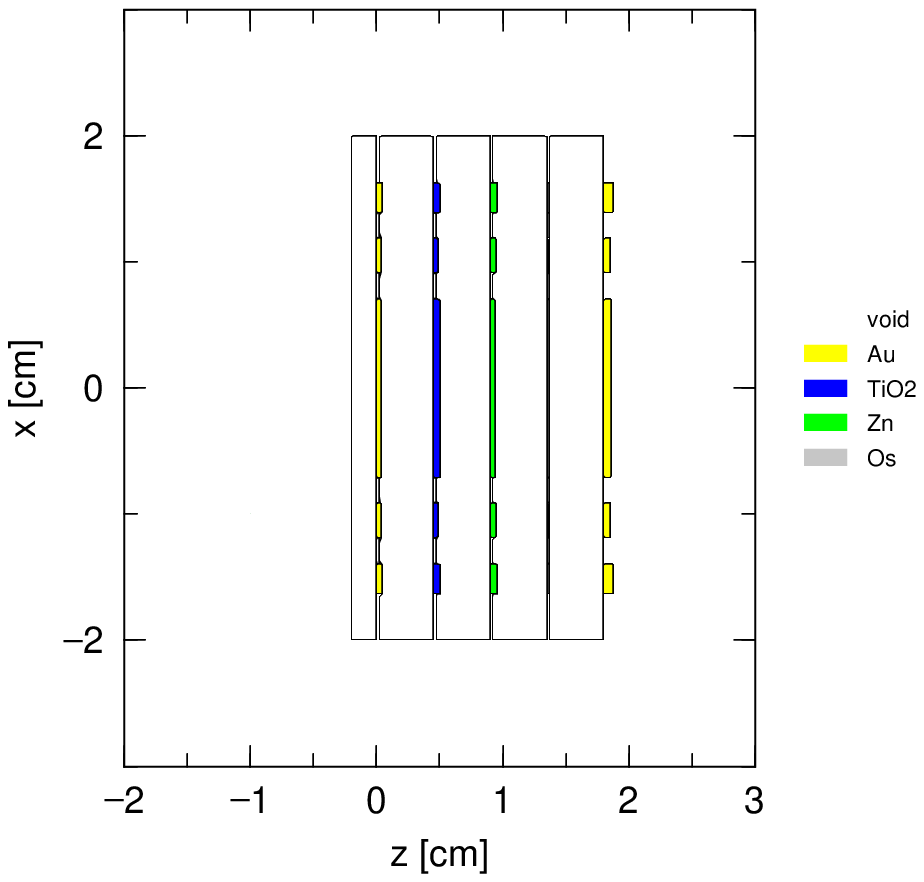}
      \color{red}
      \put(15,53){\vector(1,0){20}}
      \put(18,55){\scriptsize $\gamma$ beam}
    \end{overpic}
     \begin{overpic}[width=0.20\textwidth]{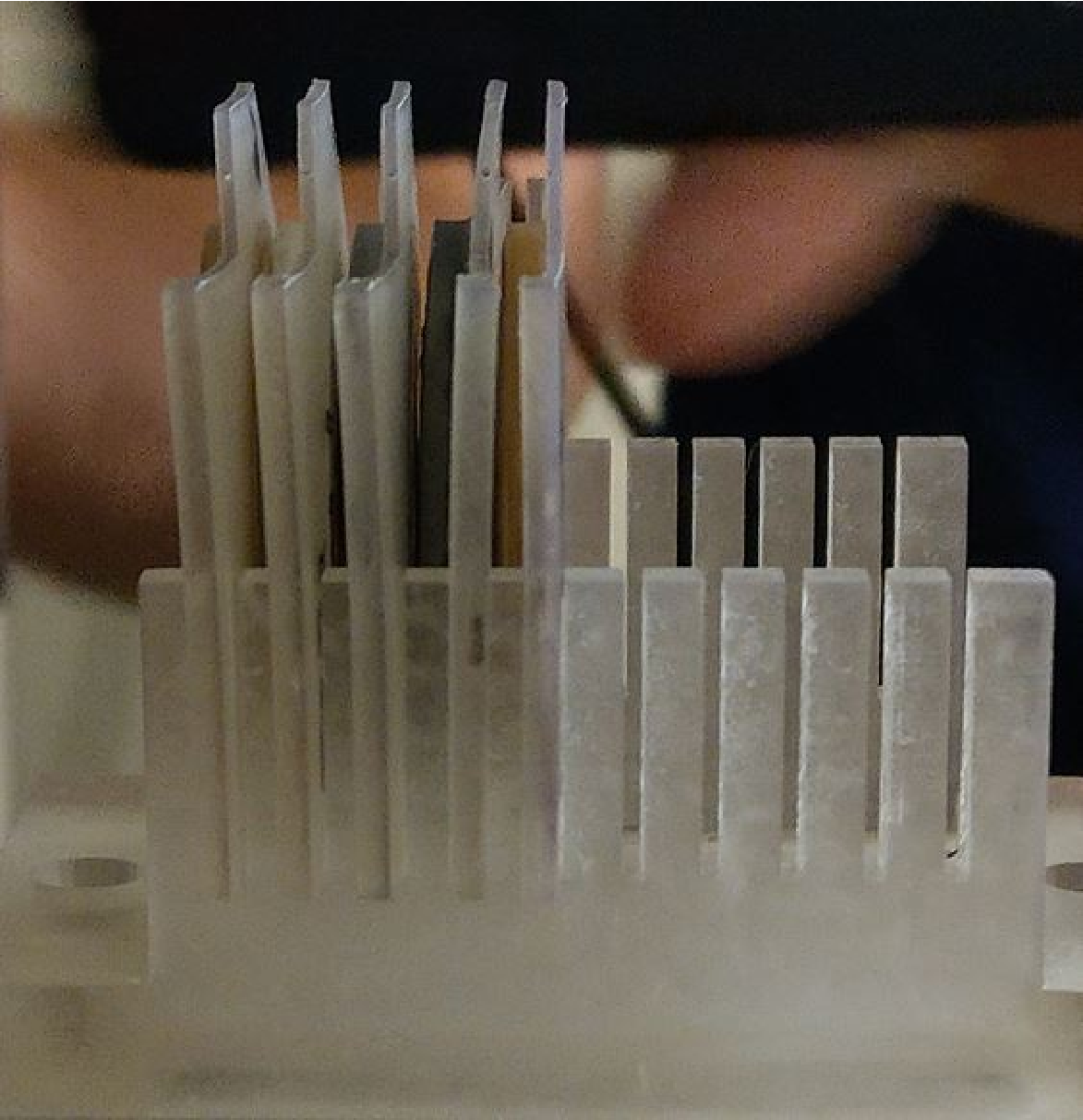}
    \end{overpic}
    \begin{overpic}[width=0.235\textwidth]{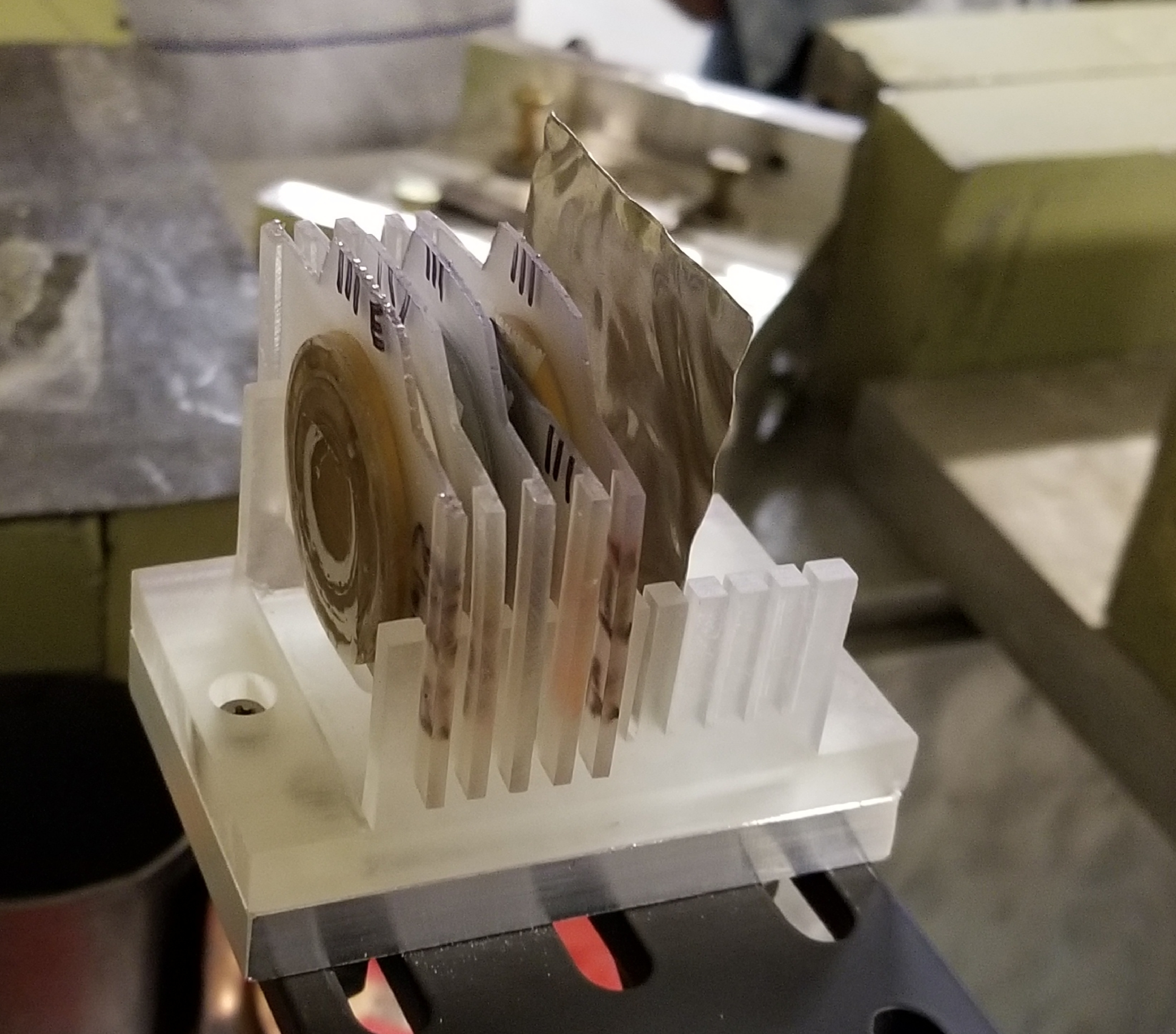}
      \color{white}
      \put(10,20){\textbf Entrance}
      \put(78,25){\textbf Exit}
      \color{white}
      \put(-85,70){\textbf Au}
      \put(-80,58){\textbf TiO$_2$}
      \put(-70,70){\textbf Zn}
      \put(-57,58){\textbf Os}
      \put(-50,70){\textbf Au}
      \color{red}
      \thicklines
      \put(-85,50){\vector(1,0){45}}
      \put(-85,43){$\gamma$ beam}

      \color{black}
      \put(-35,194){\textbf{$a$)}}
      \color{white}
      \thicklines
      \put(-82,80){\textbf{$b$)}}
      \put(5,80){\textbf{$c$)}}
    \end{overpic}
    \caption{Geometry of a target stack used in the beam flux simulations (a) and pictures of a target stack (b, c).}
    \label{fig:gem}
  \end{center}
\end{figure}

\section{Experiment}
\label{expt:tgt}
Targets were activated at seven different central photon beam energies (see Table \ref{tab:bi}).
The gamma energy at the outermost ring typically was close to the central energy of the next lower energy irradiation,
thus providing an internal consistency check for these measurements.
Using targets with three concentric segments, the cross-section measurements
for the reactions $^{197}$Au($\gamma$,n)$^{196}$Au and $^{197}$Au($\gamma$,3n)$^{194}$Au were determined
for 21 different energies.
The target stack was irradiated for 8-12 hours by a photon beam with fluxes ranging
between 10$^{8}$-10$^{9}$ $\gamma$/s, limited by a copper pre-collimator.
The boundaries of the photon beam of $\approx$ 30-40 mm in diameter were shaped
by the copper apertures inside the HI$\gamma$S aperture systems and by the copper pre-collimator.
Irradiation times, beam intensities and uncertainties are summarized in Table \ref{tab:bi}.

\begin{table}
  \caption{\label{tab:bi}Beam irradiation and intensities with associated one-sigma uncertainties.}
  \begin{ruledtabular}
    \begin{tabular}{r | r r r}
      $E_{\gamma}$      &       Irradiation time        &       Intensity       &       Unc.   \\
                MeV     &       hr      &       $\times 10^{8}$ $\gamma$/s      &       $\%$    \\
          \hline
          \hline
          14      &       11.49   &       8.01    &       8.2      \\
          16      &       10.72   &       8.80    &       8.2     \\
          18      &       8.17    &       9.27    &       8.2     \\
          22      &       10.73   &       2.48    &       8.2      \\
          24      &       11.20   &       2.70    &       8.2     \\
          27      &       11.73   &       2.51    &       8.2     \\
          31      &       9.12    &       2.06    &       8.2     \\
     \end{tabular}
  \end{ruledtabular}
\end{table}

\subsection{Target stack}
The target stack was composed of layers of Au, TiO$_{2}$, Zn, Os and Au  
as illustrated in Figure \ref{fig:gem} and listed in Table \ref{tab:tgt1}.
The target segments were mounted on 3D-printed acrylic with cavities made for receiving powdered target material.
The cavities were filled with a known mass of the target material and then sealed
by adding and curing 3D printing gel on top of this target material.
The effects of photon attenuation and scattering by the low-acrylic target mounts were negligible.
The thickness of the targets varied across the numerous ones made for this work (see Table \ref{tab:tgt1}).
The front and back Au targets were arranged in this manner 
in anticipation of using these to normalize the beam flux and calibrate the beam energy throughout the stack.
Placed front to back in the stack, relative to the incident beam, and between the front and back Au targets,  were TiO$_{2}$, Zn and Os ones.
The beam flux and scattered gamma contribution to the cross section were calculated.
There were 3 sets of targets to enable reuse after sufficient decay of the activations in subsequent runs separated by several months. 

  \begin{table}
  \caption{\label{tab:tgt1}Target stacks and thicknesses}
  \begin{ruledtabular}
    \begin{tabular}{r r r | r r r}
      & & & $E^{c}_{\gamma}$ (MeV)\\
      &  & & 14, 27 & 16, 24 & 18, 22, 31 \\
      &  & & thick.  & \\
       & tgt. &seg. &(g/cm$^{2}$) & (g/cm$^{2}$) & (g/cm$^{2}$)  \\
       \hline
       \hline
       1        &       $^{197}$Au      &       1       &       0.681   &       0.713   &       0.763   \\
                &               &       2       &       0.931   &       0.933   &       0.655   \\
                &               &       3       &       1.033   &       0.935   &       0.869   \\
                    \hline
      2 &       TiO$_2$    &       1       &       0.228   &       0.241   &       0.235   \\
              &         &       2       &       0.205   &       0.194   &       0.163   \\
              &         &       3       &       0.205   &       0.220   &       0.221   \\
                    \hline
      3 &       Zn      &       1       &       0.501&  0.350   &       0.308   \\
              &         &       2       &       0.392   &       0.292   &       0.363   \\
              &         &       3       &       0.351   &       0.339   &       0.406   \\
                    \hline
      4 &       Os      &       1       &       0.304   &       0.293   &       0.358   \\
              &         &       2       &       0.257   &       0.242   &       0.220   \\
              &         &       3       &       0.614   &       0.553   &       0.362   \\
                    \hline
      5 &       $^{197}$Au      &       1       &       1.209   &       1.299   &       1.115   \\
              &         &       2       &       0.834   &       1.195   &       1.049   \\
              &         &       3       &       1.304   &       1.395   &       1.471   \\
     \end{tabular}
  \end{ruledtabular}
\end{table}

\subsection{Photon beam flux}
\label{expt:pbf}
\begin{figure}
  \begin{center}
    \includegraphics[width=0.48\textwidth]{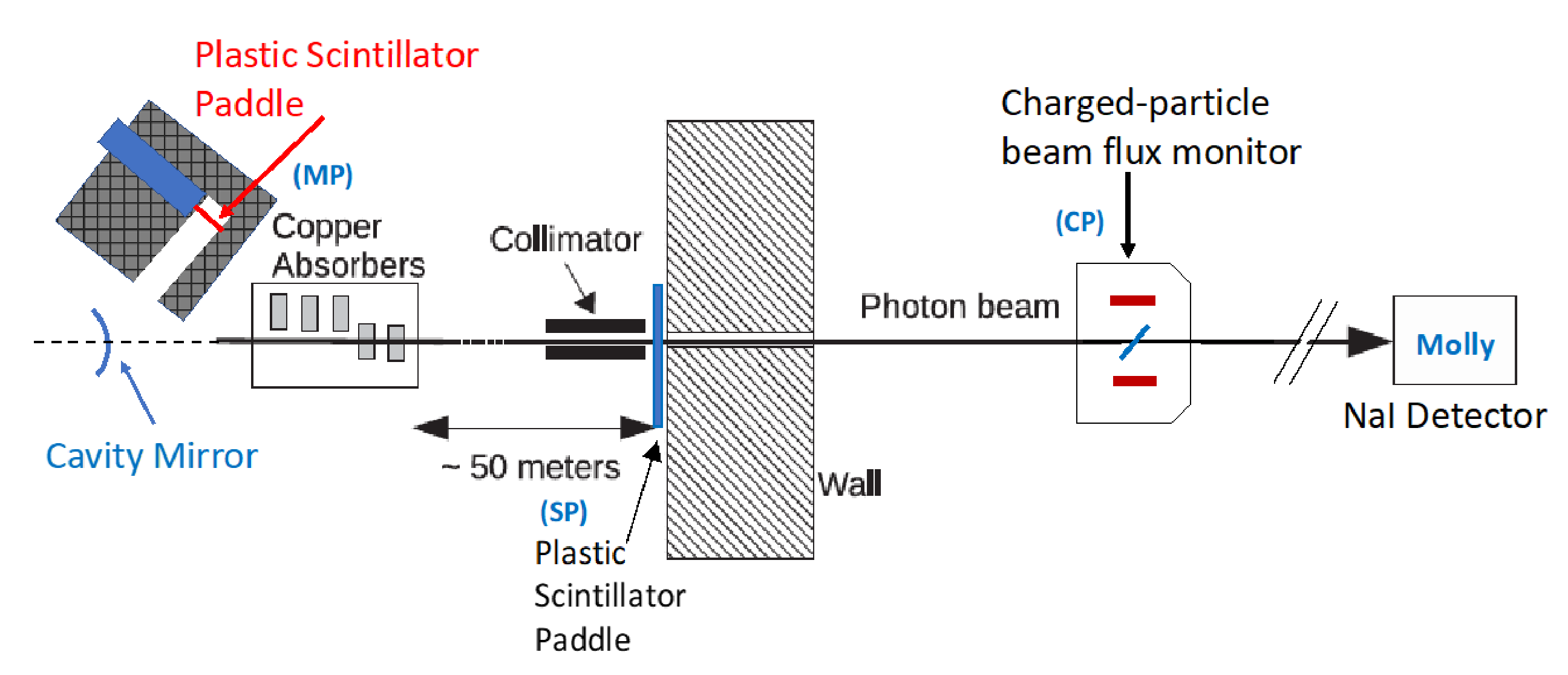}
    \caption{Schematic diagram of the experimental arrangement used to measure the $\gamma$-ray beam flux
      during target irradiation and to determine the detection efficiency of the beam flux monitors.
      The beam flux is measured during target irradiation by the detectors MP, SP and CP shown in the diagram.
      This figure is adapted from reference \cite{bf3} (see text for details).}
      \label{fig:schem}
  \end{center}
\end{figure}

\begin{figure}
  \begin{center}
    \includegraphics[width=0.45\textwidth]{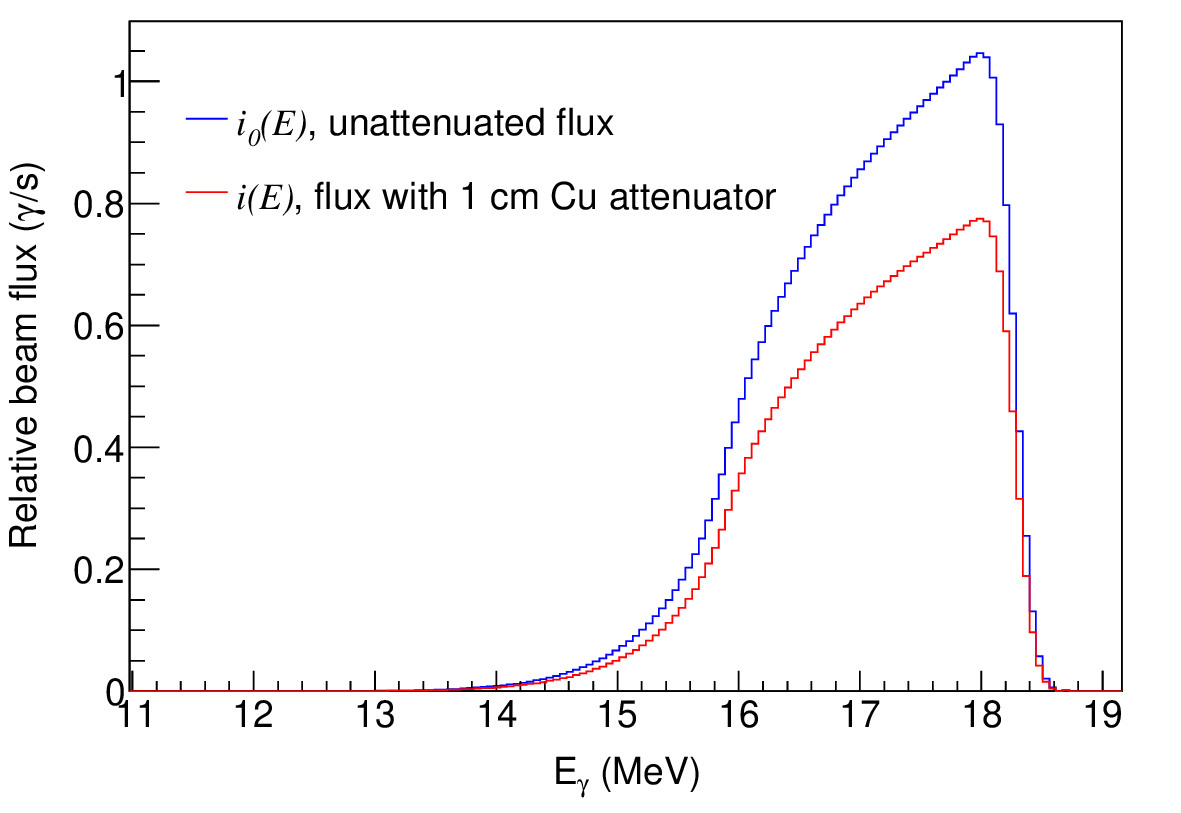}
    \caption{Plot of the simulated $i_{0}(E)$ and $i(E)$ fluxes for $E_{peak}$ = 18.0 MeV.
      The $i_{0}(E)$ beam energy profile was simulated
      by the HI$\gamma$S accelerator group.  This simulation was performed with an electron beam energy
      in the storage ring of $E_{e}$ = 675 MeV, free electron laser photon wavelength $\lambda$= 460 nm,
      a circular collimator aperture of 1.25-in diameter.
      Nominal values for the electron beam emittance parameters were used
      (see Table 2 in H. R. Weller et al. \cite{int23}).
      The distributions are plotted in 54.9 keV wide energy bins.
      The unattenuated distribution is normalized such that integral of the distribution is equal to 1.0. }
      \label{fig:relbf}
  \end{center}
\end{figure}

\begin{figure}
  \begin{center}
    \includegraphics[width=0.48\textwidth]{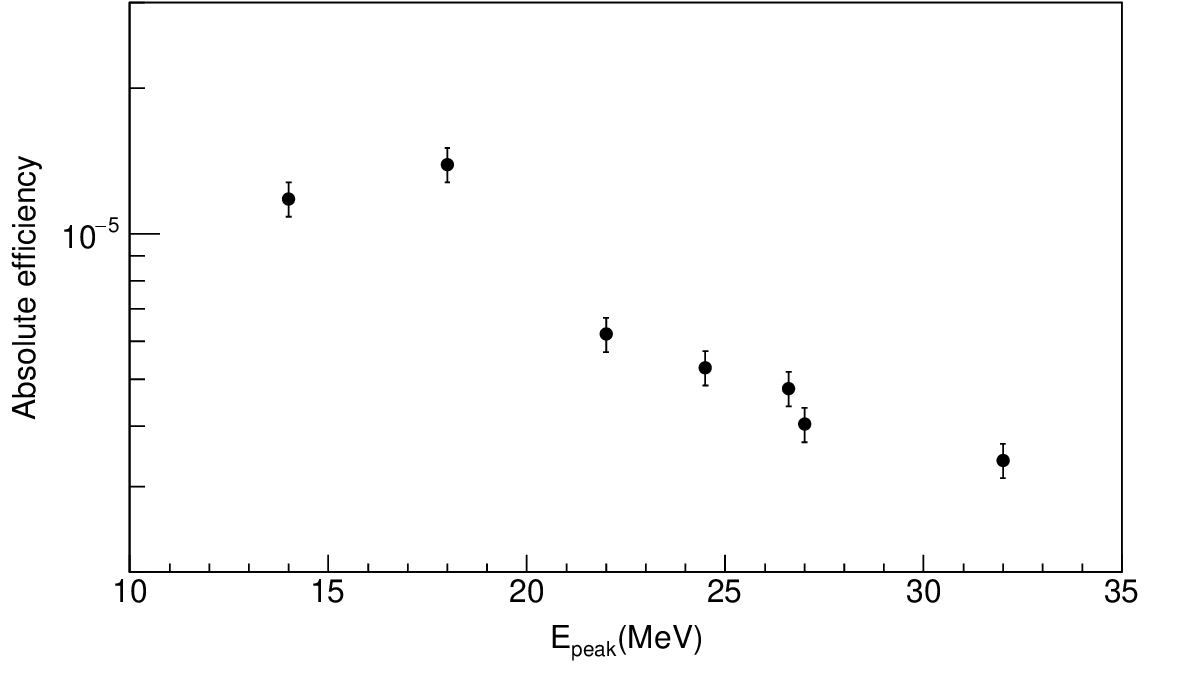}
    \caption{Plot of the measured efficiency of the MP detector as a function of $\gamma$-ray beam energy, $E_{peak}$.
      The error bars represent the combined statistical and systematic uncertainties.}
      \label{fig:mpeff}
  \end{center}
\end{figure}

A significant portion of the overall systematic uncertainty in the cross sections measured in this work is
due to the errors in measuring the absolute $\gamma$-ray beam flux.  A schematic diagram of the system
used for this purpose during target irradiation is given in Figure \ref{fig:schem}.
The beam flux was measured using three detectors during target irradiation:
1) a shielded plastic scintillator paddle which detects $\gamma$ rays and recoil electrons from Compton scattering of
the beam passing through the mirror of the optical cavity (labeled MP in Figure \ref{fig:schem}),
2) a thin plastic scintillator centered in the beam path just after the collimator
(labeled SP in Figure \ref{fig:schem}),
and 3) a silicon detector which measures charged particles emitted from photon-induced reactions
in a thin CD$_2$ foil inside an evacuated scattering chamber (labeled CP in Figure \ref{fig:schem}).
The MP detector monitors a portion of the full $\gamma$-ray beam intensity passing through the cavity mirror.
The absolute detection efficiency of MP for measuring the beam passing through the collimator is determined
using Molly (see Figure \ref{fig:schem}).  
The Molly detector is described earlier.  Photon interactions in Molly are modeled with GEANT-4.
According to these GEANT-4 simulations, the full-energy peak efficiency of Molly is greater than 98$\%$ at energies above 10 MeV. 
Molly can be positioned in the beam path, and a system of high-precision copper absorbers can be inserted
into the beam to attenuate the incident flux on Molly to keep the data acquisition system deadtime
lower than 40$\%$ and the pulse pileup rate of the detector signal below 10$\%$.
The efficiencies of the other flux monitors, SP and CP, are calibrated relative to MP using the full beam intensity.
In this arrangement, the ratios of the rates in SP and CP to that in MP are computed to monitor the stability
of the fraction of the total $\gamma$-ray flux that is transmitted through the collimator.
Over the duration of each run,  
the efficiency of MP is determined using the following relationships, 
\begin{equation}
    \label{eq:mpe}
  \epsilon_{mp}(E_{peak}) = \dfrac{I_{0}(E_{peak})}{R_{mp}(E_{peak})}
\end{equation}
where $I_{0}(E_{peak)}=I(E_{peak})e^{\langle\mu\rangle\rho x}$.
The $\langle\mu\rangle$ symbol is the photon attenuation coefficient (cm$^{2}$/g) averaged over the beam intensity profile,
$\rho$ is the density of the copper attenuators (8.96 g/cm$^{3}$), and x is the length in cm of the copper attenuator
inserted in the beam path to reduce the flux incident on Molly, while $R_{mp}$ is the net counting rate in the MP paddle.
The symbol $E_{peak}$ denotes
the $\gamma$-ray energy where the $i_{0}(E)$ distribution has a maximum value, see Figure \ref{fig:relbf}. 
The value of $\langle\mu\rangle$ was calculated using the photon attenuation coefficients
as a function of energy, $\mu$(E),
obtained from NIST \citep{bf1,bf2}, and the beam intensity profile simulated as described
in \cite{int23}.  An example of $i_{0}(E)$ and $i(E)$, the simulated unattenuated and attenuated beam intensity profiles,
used to calculate $\langle\mu\rangle$ for the beam energy distribution with
the peak beam energy, $E_{peak}$ = 18.0 MeV, is plotted in Figure \ref{fig:relbf}.  
The $E_{peak}$ value is the $\gamma$-ray energy, E, where the $i_{0}(E)$ distribution has a maximum value.
The $i_{0}(E)$ distribution is normalized such that the integration over the distribution is equal to 1.0.
The main systematic uncertainty in determining $I_{0}(E_{peak})$ in the above equation is due to the error
in the value of the attenuation $\langle\mu\rangle$.
The relative uncertainty in the measured $I(E_{peak})$ is less than 1$\%$ and is not included
in the derivation of $\Delta I_{0}$.  The relative uncertainty in $I_{0}$ is,
\begin{equation}
  \dfrac{\Delta I_{0}}{I_{0}} = \left(\dfrac{\Delta \langle\mu\rangle}{\langle\mu\rangle}\right)\rho \langle\mu\rangle x
\end{equation}
The calculated $\langle\mu\rangle$ value was benchmarked at several beam energies with in-situ measurements.
The average variance between the calculated and measured $\langle\mu\rangle$ values was $\pm$ 0.7$\%$.
We use this variance as an estimate of the uncertainty in the calculated $\langle\mu\rangle$ value.
In turn, the uncertainty in $\langle\mu\rangle$ was used to compute the average error
in the calculated $I_{0}(E_{peak})$ intensity.
For these measurements, the relative systematic uncertainty in the $\gamma$-ray beam flux is
$\Delta I_{0}/I_{0}$ = $\pm$(0.007)(8.96 g/cm$^{3}$)(0.033 cm$^2$/g)(39 cm) = $\pm$0.081 = $\pm$8.1$\%$.
The relative uncertainty in $I_{0}$ transfers to the uncertainty in measured MP efficiency, see equation \ref{eq:mpe}.
The statistical uncertainty in $R_{mp}(E_{peak})$  is less than $\pm$1$\%$.
The efficiency of MP as a function of $E_{peak}$ is displayed in Figure \ref{fig:mpeff}.
The error bars are the combined statistical and systematic uncertainties.  

\subsection{The activated target $\gamma$-ray decay spectra}

\begin{figure}
  \begin{center}
    \begin{overpic}[width=0.48\textwidth]{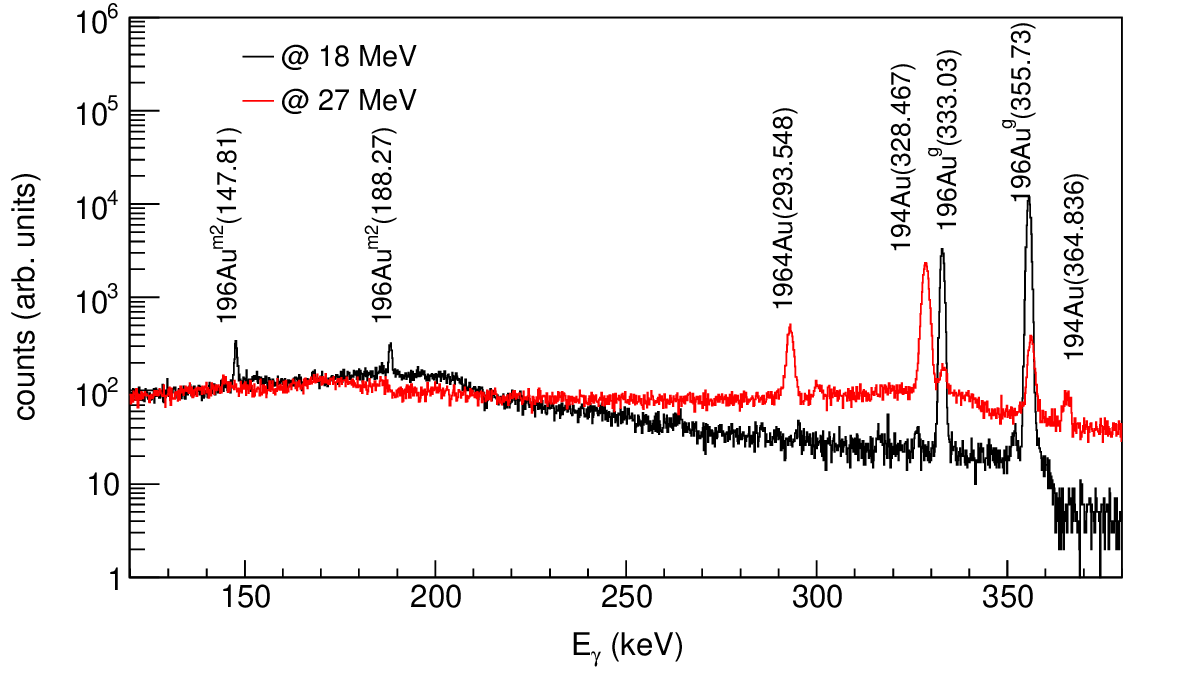}
    \end{overpic}
    \begin{overpic}[width=0.48\textwidth]{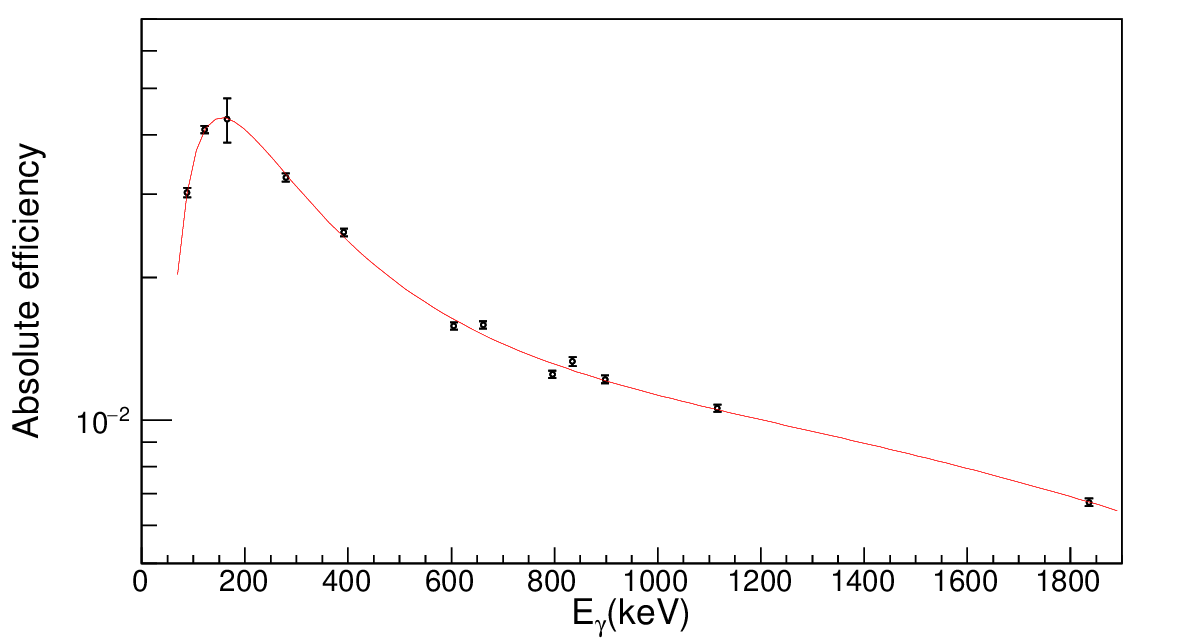}
      \put(14,107){\textbf{$a$)}}
      \put(14,47){\textbf{$b$)}}
    \end{overpic}
    \caption{a) the $\gamma$-ray spectra for the $^{197}$Au($\gamma$,xn) reactions at the photon energy of 18 MeV (black histogram)
      and 27 MeV (red histogram). b) absolute HPGe detector efficiency from the mixed calibration source. }
    \label{fig:spectra}
  \end{center}
\end{figure}

After activation, $\gamma$-ray spectra were measured with 12 High Purity Germanium (HPGe) detectors, each with a co-axial cylindrical crystal
positioned  with its front face 5 cm from the activated target.
The measurements started approximately 30 minutes after the end of bombardment and lasted long enough
to obtain sufficient $\gamma$ counts for the detection of the second isomeric state of $^{196}$Au
with the half-life of 9.6 hours. 
Typical HPGe detector efficiency shown in Figure \ref{fig:spectra} was calibrated with a mixed calibration source, and the detector efficiencies for the gold isotopes were determined
by the formula \cite{exp1}.
\begin{equation}
  \small
  \epsilon = e^{ [ (A+Bx+Cx^{2})^{-G} + (D+Ey+Fy^{2})\cdot(-G) ]^{-1/G} }
\end{equation}
where x and y are  ln$(E_{\gamma}/100)$ and ln$(E_{\gamma}$/1000), respectively, E$_{\gamma}$
is the $\gamma$-ray energy from selected nuclei.
The absolute detection efficiency at 333 keV for the $^{196}$Au transition is 1-3$\%$, 
and the corresponding uncertainty, ${\Delta \epsilon}/{\epsilon}$  is $\approx$ 2$\%$. 
During the measurements, the dead time was typically less than 2$\%$.
The spectra at the photon beam energies of 18 and 27 MeV are presented in Figure \ref{fig:spectra}.
The $\gamma$-ray peaks of 147.81, 188.27 keV (second isomeric state of $^{196}$Au), 333.03, 355.73 keV (ground state of $^{196}$Au) and
293.55, 328.47 keV ($^{194}$Au) are visible in Figure \ref{fig:spectra}.

\section{Cross-section analysis}
\label{xs:result}
The photonuclear cross sections were determined using equations \ref{eq:4} and \ref{eq:5}.
In addition, the detector efficiency, detector dead time, $\gamma$ branching ratio,
$\gamma$ beam attenuation and self-absorption effects are taken into account.
With these corrections, the cross sections can be expressed as,
\begin{equation}
  \begin{split}
    \sigma_{01}& = \dfrac{c_{all} \cdot \lambda_{0}(N_{0}^{obs}- \beta -  \gamma)}{  N_{t} I_{0} (1-e^{-\lambda_{0} t_{0}})e^{-\lambda_{0} t_{c}}(1-e^{-\lambda_{0}t_{m}})}\\
    \sigma_{2}& = \frac{ c_{all} \cdot  \lambda_{2} \cdot N_{2}^{obs}}{  N_{t} I_{0}(1-e^{-\lambda_{2}t_{0}})e^{-\lambda_{2} t_{c}}(1-e^{-\lambda_3 t_{m}})},
    \label{eq:6}
  \end{split}
\end{equation}
where $\sigma_{01}$ and $\sigma_{2}$ are the cross sections for the ground + 1$^{st}$ isomeric state
and 2$^{nd}$ isomeric state, respectively.
$N^{obs}$ stands for the detected $\gamma$ counts, $I_{0}$ and $\epsilon$ are the beam flux and the detector efficiency,
respectively.
The total correction factor $c_{all}$ above is expressed
as $c_{all}$ = $c_{\epsilon} \cdot c_{DT} \cdot c_{br} \cdot c_{ba} \cdot c_{sab}$.
The $c_{\epsilon}$, $c_{DT}$, $c_{br}$, $c_{ba}$ and $c_{sab}$ symbols are corrections for the detector efficiency,
detector dead time during the measurements,
$\gamma$-ray branching ratio, $\gamma$-ray beam attenuation and self-absorption by passing through the targets, respectively.
Correction of the primary photon beam flux due to attenuation,
and of the emitted $\gamma$-ray yield from Au isotopes due to self-absorption are taken into account by,
\begin{equation}
  C_{ba} = \dfrac{1-e^{-\mu x_{0}}}{\mu x_{0}},  C_{sab} = \dfrac{\mu x_{0}}{1-e^{-\mu x_{0}}}, 
\end{equation}

\begin{figure*}
  \begin{center}
    \includegraphics[width=0.92\textwidth]{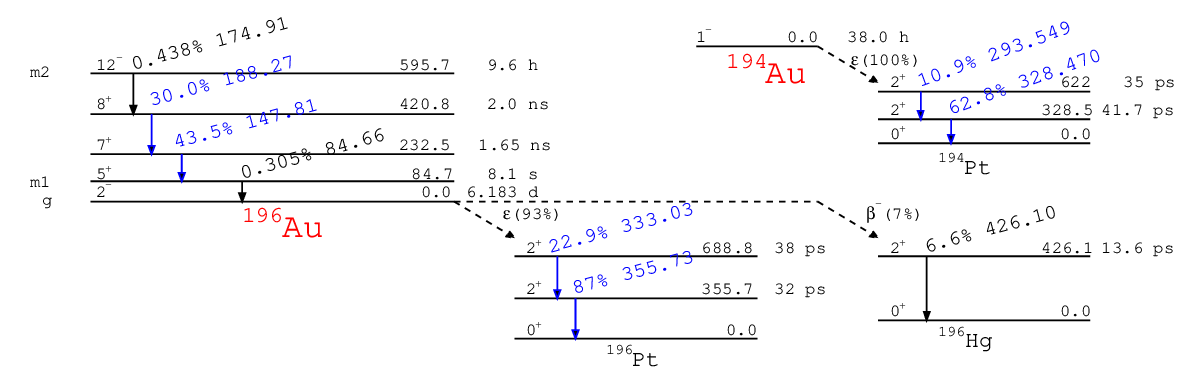}
     \caption{Simplified decay schemes for the $^{196}$Au and $^{194}$Au nuclei. The selected $\gamma$-ray lines (blue) were used in the analysis.
      The decay $\gamma$ energies and intensities are adopted from the National Nuclear Data Center (NNDC) \cite{Au1}. }
    \label{fig:ds}
  \end{center}
\end{figure*}

where $\mu$ and x$_{0}$ are attenuation coefficients for the beam and emitted $\gamma$-ray and the total length of the target, respectively.  
The statistical error was calculated by the RADWARE software \cite{exp1} and
three systematic errors for the $\gamma$-ray branching ratio, detector efficiency and beam flux were determined.
 Therefore, the total experimental error can be obtained by,
\begin{equation}
  \Delta \sigma = \sigma \sqrt{(\dfrac{\Delta N^{obs}}{N^{obs}})^{2} + (\dfrac{\Delta I_{0}}{I_{0}})^{2} +(\dfrac{\Delta \epsilon}{\epsilon})^{2}+(\dfrac{\Delta I_{br}}{I_{br}})^{2}},
\end{equation}
where $\Delta N^{obs}$,  $\Delta I_{0}$,  $\Delta \epsilon$ and  $\Delta I_{br}$ are uncertainties from the number of nuclei measured by $\gamma$ counts,
photon beam flux, HPGe detector efficiency and the $\gamma$ branching ratio of the selected nuclei, respectively.
The uncertainties of the $\gamma$ branching ratios are given in Table \ref{tab:2}.

After activation by the photon beam, the segments in the target stacks were separated and counted individually with HPGe detectors.
The perceptible Au radioisotopes produced via $^{197}$Au($\gamma$,xn) were $^{196}$Au and $^{194}$Au.
The $^{196}$Au nucleus was observed to produce two isomeric states (m$_1$= 5$^+$, m$_2$= 12$^-$) and the ground state (2$^-$).
The simplified decay schemes for the $^{196}$Au and $^{194}$Au nuclei are illustrated in Figure \ref{fig:ds}.
The half-life of the $m_{1}$, $m_{2}$ and ground states of $^{196}$Au are 8.1 s, 9.6 h and 6.17 d, respectively.
The m2 and ground states are measurable and distinguishable by counting the $\gamma$ rays from the gold targets
within a few hours from the end of each irradiation period. The m$_1$ isomeric state,
with its short half-life of only 8.1 s, has its yield included in the $\beta^{-}$ decay of the ground state.
The 147.81- and 188.24-keV $\gamma$-ray lines shown in Figure \ref{fig:ds} and the peaks in Figure \ref{fig:spectra}
are used to determine
the yield of the m$_2$ isomer, and the 333.03-, 328.47-keV $\gamma$-ray transitions
following $\beta^{+}$ emission from the ground state of $^{196}$Au are used to calculate
the cross section for the sum of the ground state and m1 isomer, respectively.
The ground state (1$^-$) of $^{194}$Au has a 38.02 h half-life and decays by $\beta^+$ emission to $^{194}$Pt.
The 293.55 and 328.47 keV $\gamma$-ray peaks were used to determine the cross section of the $^{197}$Au($\gamma$,3n)$^{194}$Au reaction.
The decay information for the $^{196}$Au and $^{194}$Au nuclei, along with uncertainties, is listed in Table \ref{tab:2}.

\begin{table}
  \caption{\label{tab:2}Decay information for the measured Au nuclei. Data are taken from the National Nuclear Data Center (NNDC) \cite{Au1}.}
  \begin{ruledtabular}
    \begin{tabular}{c |r r r r}
      nucl. & decay mode & half-life & $\gamma$ energy  & $\gamma$ branching ratio \\
        &  &  & (keV) &  ($\%$) \\
      \hline
      \hline
     $^{196}$Au$^{g}$ & $\epsilon$ 93.0 $\%$&6.1669 (6) d   & 333.03 & 22.9 (9) \\
                        &         &       & 355.73 & 87 (3)\\
     $^{196}$Au$^{m2}$&IT 100 $\%$ & 9.6 (1) h  &147.81  &  43.5 (15) \\
                      & &          &188.27  & 30.0 (15) \\
     $^{194}$Au$^{g}$&$\epsilon$ 100 $\%$ &  38.02 (10) h  &  293.549  & 10.9 (3) \\
                     & &                & 328.470    & 62.8 (16) \\
    \end{tabular}
  \end{ruledtabular}
\end{table}

\subsection{$^{197}$Au($\gamma$,n)$^{196}$Au$^{g+m_{1}, m_{2}}$}
\begin{figure*}
  \begin{center}
    \includegraphics[width=0.95\textwidth]{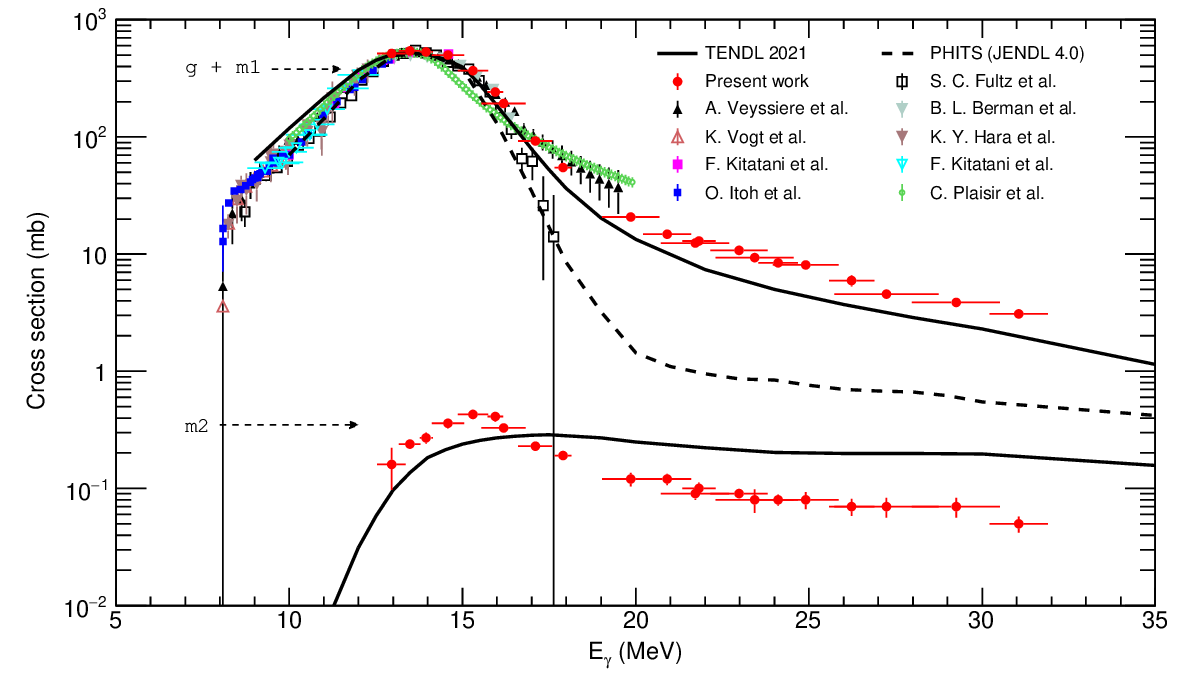}
    \caption{Excitation functions for the $^{197}$Au($\gamma$,n)$^{196}$Au$^{g+m_{1}}$  and  $^{197}$Au($\gamma$,n)$^{196}$Au$^{m_{2}}$
      reactions
      in comparison with the previous experimental data \citep{1,2,3,4,5,6,7,8,9} and the theoretical evaluations from TENDL (black line)
      and JENDL (black-dash line). Red data points indicate present measurements.
      The error bars for the present data represent the total errors, as provided in Table \ref{tab:cs196Au}.}
    \label{fig:196au_xs}
  \end{center}
\end{figure*}

The cross sections of the combined ground and first isomeric states and  second isomeric state for the photon-induced reaction on $^{197}$Au
are obtained with the emitted $\gamma$ rays of 333.03, 355.73 keV and 147.81, 188.27 keV, respectively.  
The excitation functions for the $^{197}$Au($\gamma$,n)$^{196}$Au$^{g+m_{1}}$  and  $^{197}$Au($\gamma$,n)$^{196}$Au$^{m_{2}}$ reactions
are presented in Figure \ref{fig:196au_xs} together with the previous experimental data \citep{1,2,3,4,5,6,7,8,9}
and theoretical calculations from TENDL 2021 and JENDL 4.0.
Here, the cross sections in Figure \ref{fig:196au_xs} were determined by means of weighted mean and error.
The cross sections measured in this work with statistical and total uncertainties are given in Table \ref{tab:cs196Au}.

\begin{table*}
 \footnotesize
  \caption{\label{tab:cs196Au}The photonuclear cross sections obtained for the 
    $^{197}$Au($\gamma$,n)$^{196}$Au$^{g+m_{1}}$ and $^{197}$Au($\gamma$,n)$^{196}$Au$^{m_{2}}$ reactions.
    Here, stat. and tot. stand for one-sigma statistical and total errors including the systematic ones, respectively.
    Average cross sections from the values deduced from the transition intensities are given in the columns following the individual values.}
  \begin{ruledtabular}
    \begin{tabular}{r r | r r r  rrr |rr|rrr rrr |rr }
      & nucl. &  $^{196}$Au$^{g+m_{1}}$ & & && & & &&$^{196}$Au$^{m_{2}}$ && & &&\\
      & $\gamma$ peak & 333.03 &  keV && 355.73 & keV & & && 147.81 & keV&& 188.27 &  keV&&&\\
      \hline
      $E_{\gamma}$& $\Delta {E}_{\gamma}$ & $\sigma$ & stat. & tot. & $\sigma$ & stat.  & tot. & $\bar{\sigma}$ & $\bar{\textrm{tot.}}$  & $\sigma$ &  stat.  & tot. & $\sigma$ & stat. & tot.  & $\bar{\sigma}$ & $\bar{\textrm{tot.}}$ \\
      MeV& MeV  & mb & $\%$  & $\%$  &	mb & $\%$  & $\%$   & mb & $\%$&	mb & $\%$  & $\%$   &	mb & $\%$  & $\%$ & mb & $\%$  \\
      \hline
      \hline
      12.95	&	0.41	&	518.26	&	0.20	&	9.31	&	512.33	&	0.10	&	9.12	&	514.96	&	8.72	&	0.16	&	63.6	&	64.3	&	0.16	&	50.4	&	51.4	&	0.16	&	40.27	\\
13.48	&	0.30	&	545.27	&	0.20	&	9.31	&	541.72	&	0.10	&	9.12	&	543.30	&	8.72	&	0.24	&	2.3	&	9.4	&	0.25	&	38.4	&	39.7	&	0.24	&	9.39	\\
13.96	&	0.18	&	533.03	&	0.20	&	9.31	&	530.82	&	0.10	&	9.12	&	531.81	&	8.72	&	0.28	&	16.9	&	19.2	&	0.27	&	9.2	&	13.5	&	0.27	&	10.85	\\
14.58	&	0.47	&	500.85	&	0.20	&	9.31	&	496.38	&	0.10	&	9.12	&	498.37	&	8.72	&	0.37	&	5.3	&	10.5	&	0.35	&	9.9	&	14.0	&	0.36	&	9.73	\\
15.30	&	0.43	&	370.52	&	0.20	&	9.31	&	367.09	&	0.10	&	9.12	&	368.62	&	8.72	&	0.41	&	8.3	&	12.3	&	0.44	&	5.8	&	11.4	&	0.43	&	8.67	\\
15.94	&	0.23	&	241.49	&	0.30	&	9.31	&	240.44	&	0.20	&	9.12	&	240.91	&	8.72	&	0.39	&	6.1	&	11.0	&	0.42	&	6.4	&	11.7	&	0.41	&	8.91	\\
16.18	&	0.64	&	193.69	&	0.40	&	9.32	&	192.63	&	0.20	&	9.12	&	193.10	&	8.72	&	0.30	&	7.7	&	11.9	&	0.33	&	6.6	&	11.8	&	0.33	&	9.06	\\
17.10	&	0.49	&	93.08	&	0.50	&	9.33	&	92.25	&	0.30	&	9.12	&	92.61	&	8.72	&	0.20	&	5.8	&	10.8	&	0.25	&	6.4	&	11.7	&	0.23	&	8.92	\\
17.91	&	0.24	&	54.69	&	0.60	&	9.33	&	55.28	&	0.30	&	9.12	&	55.01	&	8.72	&	0.20	&	6.1	&	11.0	&	0.19	&	5.7	&	11.3	&	0.19	&	8.59	\\
19.85	&	0.82	&	20.26	&	2.00	&	9.53	&	21.09	&	1.10	&	9.18	&	20.73	&	8.78	&	0.12	&	14.0	&	16.7	&	0.13	&	14.1	&	17.2	&	0.12	&	12.78	\\
20.90	&	0.69	&	15.27	&	4.10	&	10.16	&	14.69	&	0.70	&	9.14	&	14.86	&	8.86	&	0.12	&	9.3	&	13.0	&	0.11	&	17.4	&	20.0	&	0.12	&	11.90	\\
21.83	&	0.48	&	13.62	&	3.50	&	9.70	&	12.64	&	1.20	&	9.26	&	13.00	&	8.84	&	0.10	&	7.0	&	14.3	&	0.10	&	11.3	&	16.9	&	0.10	&	11.92	\\
21.72	&	0.99	&	12.69	&	2.70	&	9.94	&	12.23	&	1.60	&	9.20	&	12.38	&	8.86	&	0.10	&	11.0	&	11.5	&	0.09	&	13.7	&	15.0	&	0.09	&	10.34	\\
22.99	&	0.83	&	11.25	&	2.70	&	9.71	&	10.40	&	4.50	&	9.43	&	10.74	&	8.90	&	0.09	&	16.0	&	18.4	&	0.09	&	7.1	&	12.1	&	0.09	&	9.50	\\
24.12	&	0.57	&	8.28	&	4.20	&	11.78	&	8.52	&	1.40	&	9.53	&	8.46	&	9.22	&	0.09	&	43.2	&	15.4	&	0.08	&	23.6	&	10.8	&	0.08	&	10.26	\\
23.43	&	1.13	&	9.28	&	7.20	&	10.21	&	9.31	&	2.80	&	9.22	&	9.30	&	8.90	&	0.09	&	12.4	&	44.2	&	0.08	&	4.5	&	25.6	&	0.08	&	21.96	\\
24.91	&	0.95	&	7.96	&	3.30	&	9.89	&	8.16	&	1.40	&	9.23	&	8.08	&	8.86	&	0.09	&	43.1	&	44.1	&	0.08	&	15.2	&	18.1	&	0.08	&	15.94	\\
26.23	&	0.66	&	5.65	&	9.00	&	12.95	&	6.23	&	7.40	&	11.77	&	5.96	&	10.44	&	0.08	&	32.6	&	33.8	&	0.07	&	16.1	&	18.8	&	0.07	&	16.10	\\
27.24	&	1.51	&	5.22	&	2.60	&	9.67	&	4.28	&	0.50	&	9.13	&	4.56	&	8.79	&	0.08	&	19.8	&	21.8	&	0.06	&	29.7	&	31.3	&	0.07	&	18.55	\\
29.25	&	1.26	&	3.83	&	3.50	&	9.95	&	3.89	&	1.70	&	9.28	&	3.87	&	8.89	&	0.07	&	41.5	&	42.5	&	0.07	&	20.1	&	22.4	&	0.07	&	19.48	\\
31.05	&	0.84	&	3.21	&	4.40	&	10.31	&	3.02	&	1.70	&	9.28	&	3.08	&	8.94	&	0.05	&	18.3	&	20.4	&	0.05	&	17.3	&	19.9	&	0.05	&	14.87	\\
    \end{tabular}
  \end{ruledtabular}
\end{table*}

\subsection{$^{197}$Au($\gamma$,3n)$^{194}$Au}
\begin{figure}[h]
  \begin{center}
    \includegraphics[width=0.5\textwidth]{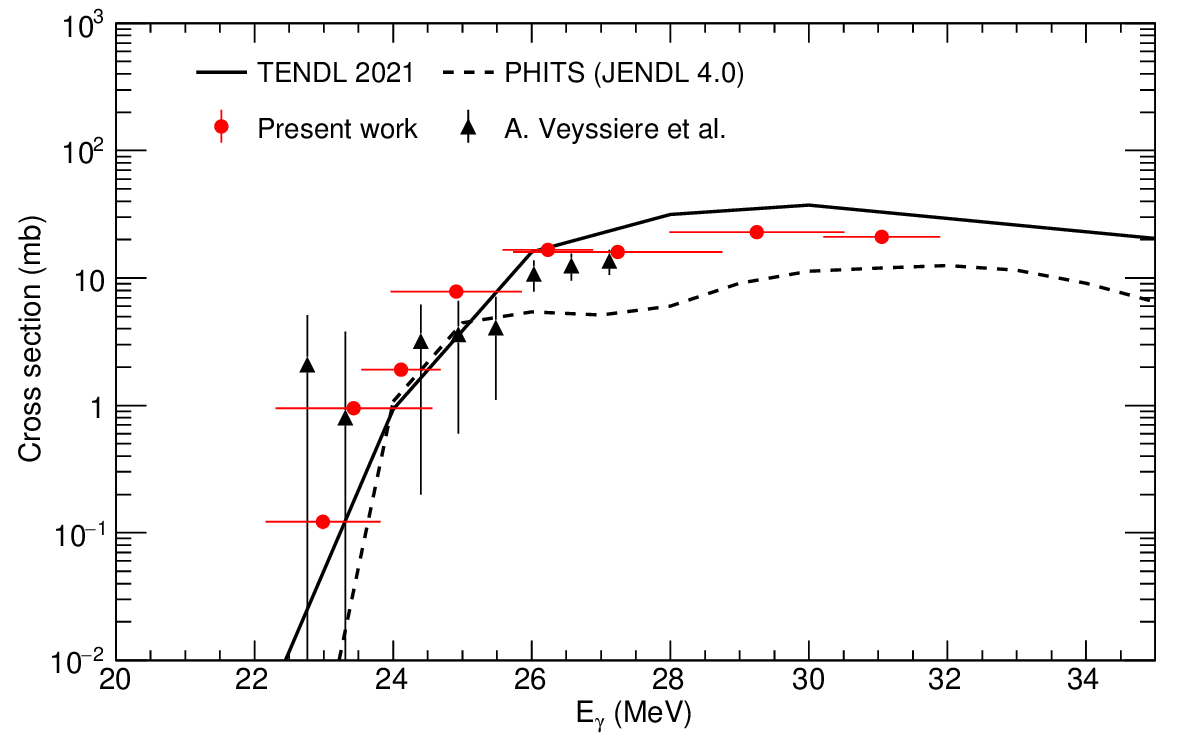}
    \caption{Excitation function for the $^{197}$Au($\gamma$,3n)$^{194}$Au reaction in comparison
      with the previous experimental data \citep{7} and the theoretical evaluation from TENDL (black line)
      and JENDL (black-dashed line). The present data points are given in red.
     The error bars for the present data represent the total errors, as provided in Table \ref{tab:cs194Au}.}
    \label{fig:194au}
  \end{center}
\end{figure}

The cross sections for the $^{197}$Au($\gamma$,3n)$^{194}$Au reaction are obtained from the emitted $\gamma$ rays of 293.55 and 328.47 keV.
The weighted mean cross sections are presented in Figure \ref{fig:194au},
together with the previous experimental data \citep{7} and theoretical calculations from TENDL 2021 and JENDL 4.0.
The cross sections measured in this work with statistical and total uncertainties are summarized in Table \ref{tab:cs194Au}.

\begin{table}[h]
   \footnotesize
  \caption{\label{tab:cs194Au}The photonuclear cross sections obtained for the 
    $^{197}$Au($\gamma$,3n)$^{194}$Au reaction.
    Here, stat. and tot. stand for one-sigma statistical and total error including the systematic ones, respectively.
  As in Table \ref{tab:cs196Au}, averaged values are given in the last columns together with the errors.}
  \begin{ruledtabular}
    \begin{tabular}{r r | r r r  rrr|rr  }
      & nucl. &  $^{194}$Au & & & & $\%$ &&\\
      & $\gamma$ peak & 293.55 &  keV && 328.46 & keV & & \\ 
      \hline
      $E_{\gamma}$& $\Delta {E}_{\gamma}$ & $\sigma$ & stat. & tot. & $\sigma$ & stat.  & tot. &  $\bar{\sigma}$  &  $\bar{\textrm{tot.}}$  \\
      MeV& MeV  & mb & $\%$  & $\%$  &  mb & $\%$  & $\%$  &  mb & $\%$ \\
      \hline
      \hline
22.99	&	0.83	&	-	&	-	&	-	&	0.12	&	5.31	&	11.5	&	0.12	&	11.5	\\
23.43	&	1.13	&	0.86	&	6.3	&	10.7	&	0.90	&	2.13	&	8.8	&	0.89	&	8.7	\\
24.12	&	0.57	&	1.83	&	7.6	&	9.0	&	2.00	&	1.22	&	8.6	&	1.93	&	8.5	\\
24.91	&	0.95	&	7.63	&	2.2	&	8.8	&	8.02	&	0.92	&	8.6	&	7.86	&	8.4	\\
26.23	&	0.66	&	16.18	&	0.9	&	8.6	&	17.02	&	1.91	&	8.8	&	16.51	&	8.4	\\
27.24	&	1.51	&	16.24	&	1.7	&	8.7	&	15.78	&	0.44	&	8.6	&	15.96	&	8.4	\\
29.25	&	1.26	&	21.25	&	1.2	&	8.6	&	24.54	&	0.39	&	8.6	&	22.86	&	8.4	\\
31.05	&	0.84	&	20.41	&	1.3	&	8.7	&	21.70	&	0.41	&	8.6	&	21.10	&	8.4	\\
    \end{tabular}
  \end{ruledtabular}
\end{table}

\section{Results and Discussion}

\label{RD:result}

This work demonstrates the successful development of a novel and highly efficient method for concurrently measuring
photonuclear reaction cross sections at multiple discrete photon energies
for various materials using a single irradiation.
As a proof-of-concept,
the excitation function of the $^{196}$Au($\gamma$,n)$^{196}$Au$^{g+m_{1},m_{2}}$ and $^{196}$Au($\gamma$,3n)$^{194}$Au reactions
were measured with this novel method utilizing target stacks
consisting of concentric ring segments bombarded
with an uncollimated beam with well-defined radial energy and flux distributions.
The measurements were performed at 21 discrete beam energies
over the incident photon beam energy range of 13-31 MeV (see Table \ref{tab:cs196Au}).
For the first time, the cross sections of the combined ground and first isomeric states at beam energies above 20 MeV and
the second isomeric state for the $^{197}$Au($\gamma$,n)$^{196}$Au reactions,
were determined via the activation method which yielded
exclusive cross sections which are unaffected by multiple neutron emission.
The method developed here based on radial target segmentation and target stacking makes
efficient use of beam time by concurrent irradiation
of multiple targets at multiple discrete $\gamma$-ray beam energies.
Our $^{197}$Au($\gamma$,n) cross-section data for the production of $^{196}$Au in the ground state are in agreement
with previous data measured using mono-energetic photon beams, but differ from data measured with a bremsstrahlung beam above 14 MeV
(see Figure \ref{fig:196au_xs} and Table \ref{tab:cs196Au}).
Our data, which were measured with a mono-energetic $\gamma$-ray beam using activation techniques,
are in good agreement with data measured with a similar beam but using neutron counting techniques.
This result suggests that the standing discrepancies between data are most likely associated with beam type
and not experimental techniques, i.e., neutron counting versus target activation.
Furthermore, the TENDL and JENDL evaluations
significantly underpredict the cross sections at energies above 18 MeV, suggesting a need to update these evaluations.
Our data for the $^{197}$Au($\gamma$,3n) reaction are given in Table \ref{tab:cs194Au}
and plotted in Figure \ref{fig:194au} in comparison with previous data
and model calculations. As can be inferred from Figure \ref{fig:194au}, additional data at energies above 30 MeV are desirable
to better estimate the predictive powers of the TENDL and JENDL evaluations.
As discussed in Section \ref{introd:int}, the ($\gamma$,p) reaction
is the primary pathway for producing high specific-activity radioisotopes using bremsstrahlung $\gamma$ rays
generated by electron LINACs.
Our cross-section analysis can be extended to the remaining targets (TiO$_2$, Zn, and Os) to obtain ($\gamma$,p)
reaction cross sections.
The main reactions of interest for the remaining targets are $^{48}$Ti($\gamma$,p)$^{47}$Sc, $^{68}$Zn($\gamma$,p)$^{67}$Cu,
and $^{189}$Os($\gamma$,p)$^{188}$Re. The data obtained for these reactions will be reported in future publications.

\begin{figure}
  \begin{center}
    \begin{overpic}[width=0.49\textwidth]{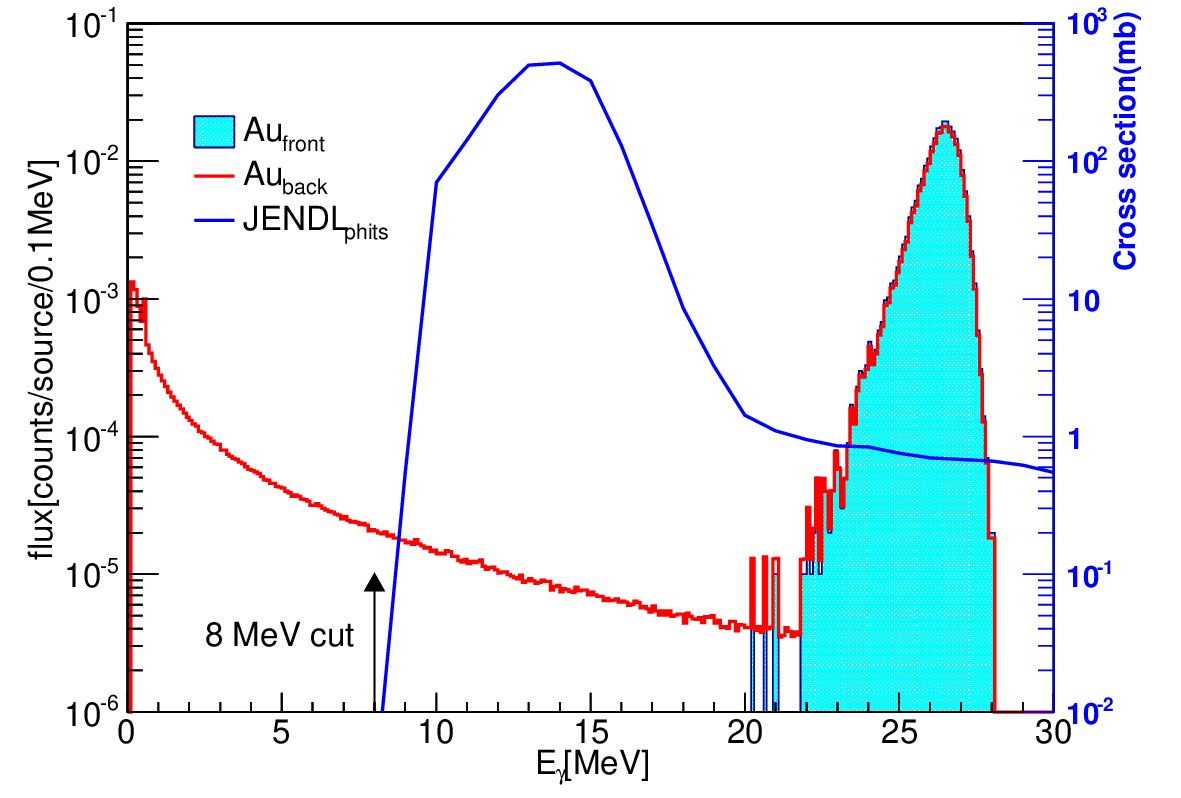}
    \end{overpic}
    \begin{overpic}[width=0.45\textwidth]{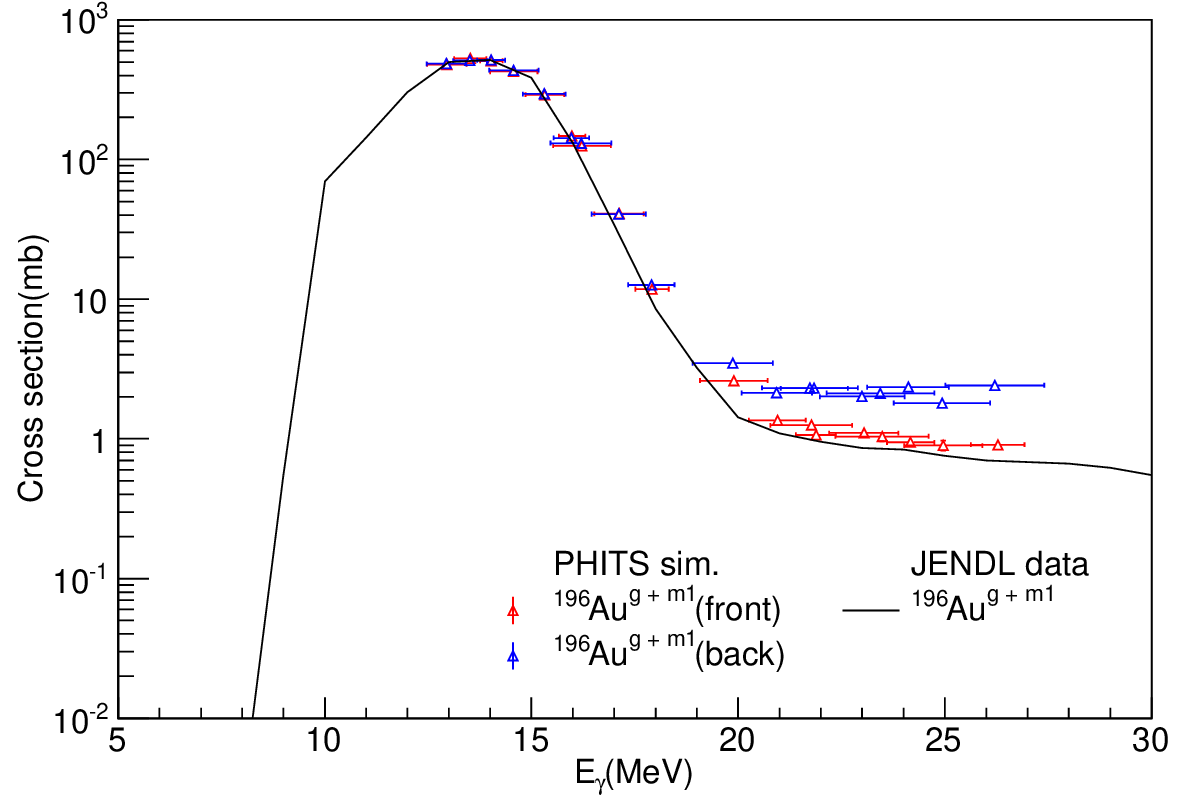}
      \put(15,134){$a$)}
      \put(15,61){$b$)}
    \end{overpic}
    \caption{The PHITS simulation results:
      a) The photon flux as a function of energy before the first Au and before the back Au target,
      overlaid with the cross sections for the $^{197}$Au($\gamma$,n) reaction,
      b) The effective cross sections for the
      front and back Au targets extracted from PHITS simulations using the JENDL photo-nuclear libraries.
      Red and blue markers indicate the front and back Au targets, respectively.}
    \label{fig:scat}
  \end{center}
\end{figure}

\subsection{Incident photon scattering effect}
Having accurate values for the beam flux at each segment of each target
is essential in determining the corresponding reaction cross sections.
The fractional photon beam fluxes at each target segment were obtained from simulations, which were benchmarked
with measurements as described above.
With these incident beam fluxes on the segments of the front target,
the beam intensity at each target was calculated using the beam transport simulation performed
by the PHITS Monte-Carlo code \cite{beam3}.
The geometry illustrated in Figure \ref{fig:gem} was implemented in the simulations.
The targets in the stack were Au, TiO$_2$, Zn, Os and Au. Successive targets were separated by 0.45 cm.
The detailed information about the segments of the actual target stack can be found in Section \ref{expt:tgt}.
A total of 15 segments in the target stack (3 segments per target) were exposed to the photon beam at once.
The beam flux was simulated for each irradiation (See Table \ref{tab:fr}).
In the energy range of our measurements, Compton scattering and pair production dominate the beam attenuation and scattering
as it passes through the stack.
The attenuation of the gamma beam is simulated in detail using the PHITS code
with the overall attenuation predicted to be $\approx$ 10$\%$.
The central beam energy, one-sigma radial energy spread, and fractional beam flux on each segment of the stack are given
in Table \ref{tab:fr}.
The overall simulation is validated using the data from the gold target at the back of the stack when compared with the data from
the one in the front.
Predictions of distortions to the photon energy spectrum due to the photons scattering
as they pass through the target stack are made using PHITS.   
Some results from the PHITS calculations are found in Figure \ref{fig:scat}.
In these calculations $E_{peak}$ is about 25 MeV, as can be inferred from the shaded peak in Figure \ref{fig:scat} a).
As the photon bunch traverses the target stack, the energy of a small fraction of the photons is degraded
due to Compton scattering from atomic electrons.  This effect creates the low-energy tail on the photon energy distribution
(red curve) displayed in Figure \ref{fig:scat} a).
Because this low-energy tail extends through the GDR region, the cross section measured
with the back Au target is higher than that measured using the front one in the beam energy range of 20-27 MeV. 
This effect increases the apparent cross section in the high-energy region by ~ 2 mb, but has a very small effect
in the giant resonance region where the cross sections are approximately 100 times larger,
as shown by the simulations plotted in Figure \ref{fig:scat} b).
Cross sections for the other targets in the stack will be corrected
for this effect in future publications.

\subsection{Beam flux correction on the third segment }
The photon beam at the outer radii (r $\approx$ 15 mm) for the third segment of each target was partially blocked 
by the copper apertures inside the vacuum chamber for the FEL aperture system of HI$\gamma$S.
The beam flux on the third segment was found to be underestimated in the simulation.
The correction factor of the beam flux on the third segment was determined to be 8$\%$.
The cross sections for the third segments in Table \ref{tab:cs196Au} and \ref{tab:cs194Au} were adjusted
to account for this effect.

\section{Conclusions}

We report cross-section measurements for the $^{197}$Au($\gamma$,n)$^{196}$Au$^{g+m1}$, $^{197}$Au($\gamma$,n)$^{196}$Au$^{m2}$,
and $^{197}$Au($\gamma$,3n)$^{194}$Au reactions at $\gamma$-ray beam energies ranging from 13 to 31 MeV.
The measurements were performed at HI$\gamma$S demonstrating the enhanced efficacy of a new sample activation method based
on irradiating targets consisting of concentric-ring segments relative to using solid targets.  In this proof-of-concept experiment,
data at three discrete $\gamma$-ray energies were collected for each target irradiation.
Our results represent the first cross-section data for the $^{197}$Au($\gamma$,n)$^{196}$Au$^{g+m1}$  reaction above 20 MeV.
As such, they enable the first data-based assessment of the predictions of the TENDL \cite{int7} and JENDL \citep{int8,int9,int10,int11}
evaluated nuclear data libraries above 20 MeV for this reaction and provide new data
for updating the reaction models used in the evaluations.  Also, our data for the $^{197}$Au($\gamma$,n)$^{196}$Au$^{m2}$ reaction,
leaving $^{196}$Au in the $J^{\pi}$ = $12^{-}$ metastable state ($T_{1/2}$ = 9.6 h), are the first, and thereby provide constraints
for the modeling of this reaction in TENDL. These new measurements have significant implications
on the use of the $^{197}$Au($\gamma$,n)$^{196}$Au$^{g+m1}$ reaction as a cross-section standard.
Our data for the $^{197}$Au($\gamma$,n)$^{196}$Au$^{g+m1}$ reaction are in good agreement with TENDL predictions and existing data
that were measured using a monoenergetic $\gamma$-ray beam and neutron counting techniques.
However, they differ by as much as 60$\%$ from data measured with a bremsstrahlung beam at energies between 13 and 16 MeV.
This observation suggests that, in the energy range of 13 – 16 MeV, preference should be given
to using cross sections for the $^{197}$Au($\gamma$,n)$^{196}$Au$^{g+m1}$ reaction that are consistent
with data measured using monoenergetic $\gamma$-ray beams.


\section{Acknowledgments}

This research was supported by the U.S. Department of Energy Isotope Program, managed by the Office of
Science for Isotope R$\&$D and Production, under
contract DE-AC02-06CH11357 (Argonne National Laboratory),
and the Department of Energy, Office of Nuclear Physics under grant numbers
DE-SC0018112, DE-SC0018325, DE-FG02-97ER41033 (TUNL) and DEFG02-97ER41041 (UNC).
The authors would like to thank the HI$\gamma$S operators and staff for their help during the experiments.
We gratefully acknowledge the computing resources provided on Bebop, a high-performance computing cluster operated by the Laboratory Computing Resource Center at Argonne National Laboratory, and the assistance of Dr. Sean Finch at TUNL in the analysis of targets used in this research.

\appendix{}

\section{Activation}
\label{appdx:act}

The reaction rate (R) is given by,
\begin{equation}
  R = \sigma \cdot N_{t}\cdot I_{0}
\label{eq:0}
\end{equation}
where $\sigma$ is the cross section, $N_{t}$ is the number of atoms in the target, and $I_{0}$ is the beam flux of incident particles.

\begin{figure}[h]
  \begin{center}
  \begin{overpic}[width=0.45\textwidth]{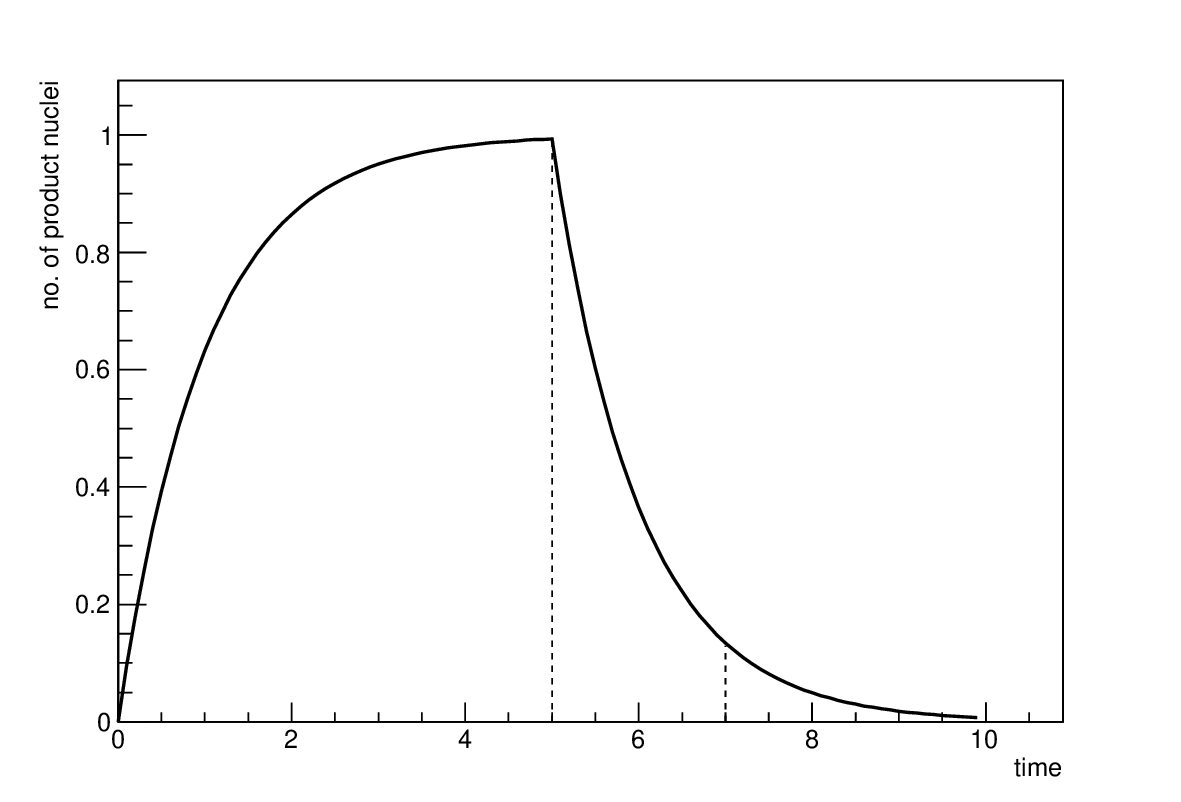}
    \tiny
    \put(10,57){\vector(1,0){37}}
    \put(47,57){\vector(-1,0){37}}
    \put(62,22){\vector(-1,0){15}}
    \put(47,22){\vector(1,0){15}}
    \color{red}
    \put(22,52){\large $t_{0}$}
    \put(55,24){\large $t_{c}$}
    \color{black}
    \put(62,7){\line(0,1){7}}
    \put(72,7){\line(0,1){7}}
    \color{blue}
    \put(72,14){\vector(-1,0){10}}
    \put(62,14){\vector(1,0){10}}
    \color{black}
    \color{red}
    \put(65,16){\large $t_{m}$}
  \end{overpic}
  \end{center}
  \caption{Number of radioactive nuclei in the targets as a function of time during and after bombardment.
    The $t_{0}$, $t_{c}$ and $t_{m}$ symbols are the irradiation, cooling and measurement time periods, respectively.}
  \label{fig:prod}
\end{figure}
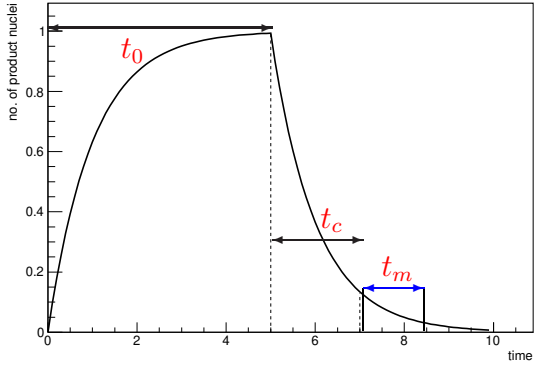

An example of a radioisotope production curve is provided in Figure \ref{fig:prod}. The isotope is being produced faster than
it can decay during the time from 0 to $t_{0}$ when the target is being irradiated.
\subsubsection{$^{197}$Au($\gamma$,n)$^{196}$Au$^{g,m_{1},m_{2}}$ reaction}
The $^{196}$Au nucleus has three different long-lived states; e.g. the ground, $1^{st}$ and $2^{nd}$ isomeric states denoted by g (6.17 d), $m_{1}$ (8.1 s) and $m_{2}$ (9.6 h).
The two isomeric levels decay 100$\%$  to the ground state.
The number of nuclei produced during the irradiation time
can be calculated using the formulae below,
\begin{equation}
  \begin{split}
  dN_{g} & = R_{g}dt - \lambda_{g}N_{g} dt + \lambda_{m1}N_{m1}dt +  \lambda_{m2}N_{m2}dt \\
  dN_{m1} & = R_{m1}dt - \lambda_{m1}N_{m1} dt \\
  dN_{m2} & = R_{m2}dt - \lambda_{m2}N_{m2} dt,
  \end{split}
  \label{eq:1}
\end{equation}

where the R's and $\lambda$'s are the reaction rates and decay constants, respectively. $N_{i}$ is the number of nuclei in the $i$th state.
The solutions for the number of nuclei in the $1^{st}$ and $2^{nd}$ isomeric states are obtained from the following equations.

\begin{equation*}
  \begin{split}
    N_{m1} & = \frac{R_{m1}}{\lambda_{m1}}(1-e^{-\lambda_{m1}t}) \approx \frac{R_{m1}}{\lambda_{m1}} \\
    N_{m2} & = \frac{R_{m2}}{\lambda_{m2}}(1-e^{-\lambda_{m2}t})\\
  \end{split}
\end{equation*}

The $N_{m1}$ number can be approximately constant, since $\lambda_{m1} \,t \gg 1$
due to the extremely short half-life compared to that of ground and $2^{nd}$ isomeric states.
This means  $N_{g} = N_{g}^{\prime} + N_{m1}$.
In other words, we can determine the cross sections for the production of $^{196}Au^{g+m_{1}}$ and $^{196}Au^{m_{2}}$, separately.
Using the above solutions and equation \ref{eq:1}, we find the solution for $N_{g+m1}$,
\begin{equation*}
  \begin{split}
  N_{g}(t)  = & \frac{R_{g+m1}}{\lambda_{g}}(1-e^{-\lambda_{g}t})\\
  &+\frac{R_{m2}}{\lambda_{g}}(1-e^{-\lambda_{g}t})
  -\frac{R_{m2}}{\lambda_{m2} - \lambda_{g}}(e^{-\lambda_{g}t}-e^{-\lambda_{m2}t}),
  \end{split}
\end{equation*}
where $R_{g+m1} = R_{g} + R_{m1}$.
When the irradiation is finished, the nuclei will decay with no production term.
The number of nuclei decreases with the respective half-lives,

\begin{equation}
  \begin{split}
  dN_{g} & =-\lambda_{g}N_{g}dt +\lambda_{m2}N_{m2}dt \\
  dN_{m2} & =-\lambda_{m2}N_{m2}dt
  \end{split}
  \label{eq:2}
\end{equation}
The number of nuclei can be calculated from integrating equation \ref{eq:2},
\begin{equation*}
  \begin{split}
  N_{m2}(t) & = N_{m2}^{0} e^{-\lambda_{m2}t} \\
  N_{g}(t) & = N_{g}^{0}e^{-\lambda_{g}t}+N_{m2}^{0}\frac{\lambda_{m2}}{\lambda_{m2} - \lambda_{g}}(e^{-\lambda_{g}t}-e^{-\lambda_{m2}t})
  \end{split}
\end{equation*}

with,
\begin{equation*}
  \begin{split}
    N_{g}^{0} & = \frac{R_{g+m1}}{\lambda_{g}}(1-e^{-\lambda_{g}t_{0}})\\
    &+\frac{R_{m2}}{\lambda_{g}}(1-e^{-\lambda_{g}t_{0}})
    -\frac{R_{m2}}{\lambda_{m2} - \lambda_{g}}(e^{-\lambda_{g}t_{0}}-e^{-\lambda_{m2}t_{0}})\\
    N_{m2}^{0} & = \frac{R_{m2}}{\lambda_{m2}}(1-e^{-\lambda_{m2}t_{0}})
  \end{split}
\end{equation*}
The total number of events with measurement time($t_{m}$) will be,
\begin{equation*}
  \begin{split}
    N_{m2}^{evt}  ={}&  N_{m2}^{0} e^{-\lambda_{m2}t_{c}}(1-e^{-\lambda_{m2} t_{m}})  \\
    N_{g}^{evt} ={}&  N_{g}^{0} e^{-\lambda_{g}t_{c}} (1-e^{-\lambda_{g} t_{m}}) + N_{m2}^{0} \frac{\lambda_{m2}\lambda_{g}}{\lambda_{m2} - \lambda_{g}} \times \
\\
    & (\frac{e^{-\lambda_{g}t_{c}}}{\lambda_{g}}(1-e^{-\lambda_{g} t_{m}})-\frac{e^{-\lambda_{m2}t_{c}}}{\lambda_{m2}}(1-e^{-\lambda_{m2} t_{m}}))
  \end{split}
\end{equation*}

Using these results and the relation between the number of events and the cross section in equation \ref{eq:0},
the cross section for producing $^{196}$Au in the 2$^{nd}$ isomer state is,

\begin{equation}
   \sigma_{m2}  = \frac{\lambda_{m2} N_{m2}^{evt}}{ N_{t} I_{0}(1-e^{-\lambda_{m2}t_{0}})e^{-\lambda_{m2} t_{c}}
      (1-e^{-\lambda_{m2} t_{m}})}
  \label{eq:4}
\end{equation}

Similarly, the combined cross section for ground and $1^{st}$ isomeric state can be obtained,
\begin{equation}
  \sigma_{g+m1} = \dfrac{\lambda_{g}(N_{g}^{evt}- \beta -  \gamma)}
        { N_{t} I_{0} (1-e^{-\lambda_{g} t_{0}})e^{-\lambda_{g} t_{c}}(1-e^{-\lambda_{g}t_{m}})}
        \label{eq:5}
\end{equation}

with,
\begin{equation*}
  \footnotesize
  \begin{split}
    \alpha & = \frac{R_{m2}}{\lambda_{g}}(1-e^{-\lambda_{g}t_{0}})-\frac{R_{m2}}{\lambda_{m2} - \lambda_{g}}(e^{-\lambda_{g}t_0{}}-e^{-\lambda_{m2}t_{0}})\\
    \beta & = N_{m2}^{0} \frac{\lambda_{m2}\lambda_{g}}{\lambda_{m2} - \lambda_{g}}
    (\frac{e^{-\lambda_{g}t_{c}}}{\lambda_{g}}(1-e^{-\lambda_{g} t_{m}})-\frac{e^{-\lambda_{m2}t_{c}}}{\lambda_{m2}}(1-e^{-\lambda_{m2} t_{m}}))  \\
    \gamma & = \alpha e^{-\lambda_{g}t_{c}}(1-e^{-\lambda_{g} t_{m}})
  \end{split}
\end{equation*}

\subsubsection{$^{197}Au(\gamma,3n)^{194}Au$ reaction}
The cross section for producing $^{194}Au$ is determined in the same way as Equation \ref{eq:4},
\begin{equation}
  \small
  \begin{split}
    \sigma  = \frac{\lambda N^{obs}}{ N_{t} I_{0} (1-e^{-\lambda t_{0}})e^{-\lambda t_{c}}
      (1-e^{-\lambda t_{m}})   }
  \end{split}
  \label{eq:61}
\end{equation}

\section{Beam flux simulation results}
Beam flux simulation results for each segment and target are listed in the Table \ref{tab:fr}.

\begin{table*}
  \caption{\label{tab:fr}Beam flux simulation results}
  \begin{ruledtabular}
    \begin{tabular}{r r | r r r|rrr | rrr | rrr}
      & E$_{\gamma}^{center}$ & 14 MeV &&& 16 MeV &&& 18 MeV &&& 22 MeV\\
       seg & tgt. & $E_{\gamma}$ & $\Delta E_{\gamma}$ & flux ($\%$) & $E_{\gamma}$ & $\Delta E_{\gamma}$ & flux ($\%$) & $E_{\gamma}$ &$\Delta E_{\gamma}$ & flux ($\%$)  &   $E_{\gamma}$ & $\Delta E_{\gamma}$ & flux ($\%$)\\
      \hline
      \hline
      1 &       Au      &       13.956  &       0.181   &       23.960  &       15.944  &       0.227   &       24.400  &       17.907  &       0.243   &       24.920  &       21.828  &       0.483   &       24.680  \\
      & TiO$_{2}$       &       13.956  &       0.182   &       23.080  &       15.944  &       0.226   &       23.430  &       17.906  &       0.243   &       23.840  &       21.825  &       0.497   &       23.536  \\
      & Zn      &       13.956  &       0.181   &       22.960  &       15.944  &       0.226   &       23.300  &       17.906  &       0.243   &       23.710  &       21.824  &       0.500   &       23.404  \\
       & Os      &       13.956  &       0.182   &       22.670  &       15.944  &       0.227   &       23.090  &       17.906  &       0.243   &       23.520  &       21.823  &       0.504   &       23.204  \\
      & Au      &       13.956  &       0.182   &       22.300  &       15.944  &       0.227   &       22.720  &       17.906  &       0.243   &       23.050  &       21.822  &       0.510   &       22.709  \\
      \hline
       2 &       Au      &       13.477  &       0.303   &       23.320  &       15.303  &       0.430   &       23.180  &       17.100  &       0.486   &       23.160  &       20.903  &       0.692   &       23.285  \\
      & TiO$_{2}$       &       13.477  &       0.303   &       22.280  &       15.303  &       0.430   &       21.990  &       17.100  &       0.486   &       22.300  &       20.901  &       0.697   &       22.363  \\
      & Zn      &       13.477  &       0.303   &       22.180  &       15.302  &       0.430   &       21.890  &       17.100  &       0.486   &       22.220  &       20.901  &       0.697   &       22.276  \\
       & Os      &       13.477  &       0.303   &       21.960  &       15.302  &       0.430   &       21.730  &       17.100  &       0.486   &       22.000  &       20.900  &       0.699   &       22.053  \\
      & Au      &       13.477  &       0.303   &       21.650  &       15.302  &       0.430   &       21.440  &       17.100  &       0.486   &       21.730  &       20.900  &       0.700   &       21.763  \\
       \hline
        3 &       Au      &       12.915  &       0.414   &       19.270  &       14.581  &       0.468   &       18.830  &       16.184  &       0.637   &       18.610  &       19.854  &       0.824   &       18.871  \\
      & TiO$_{2}$       &       12.914  &       0.414   &       18.230  &       14.581  &       0.468   &       17.880  &       16.184  &       0.637   &       17.720  &       19.852  &       0.827   &       17.903  \\
      & Zn      &       12.914  &       0.414   &       18.140  &       14.581  &       0.468   &       17.780  &       16.184  &       0.637   &       17.620  &       19.852  &       0.828   &       17.807  \\
         & Os      &       12.914  &       0.414   &       17.980  &       14.581  &       0.468   &       17.630  &       16.183  &       0.637   &       17.430  &       19.851  &       0.828   &       17.609  \\
      & Au      &       12.914  &       0.415   &       17.410  &       14.580  &       0.468   &       17.100  &       16.183  &       0.638   &       17.090  &       19.851  &       0.829   &       17.235  \\
      \hline
      \hline
      & E$^{center}_{\gamma}$ & 24 MeV &&& 26 MeV &&& 31 MeV &&& \\
      seg & tgt. & $E_{\gamma}$ & $\Delta E_{\gamma}$ & flux ($\%$) & $E_{\gamma}$ & $\Delta E_{\gamma}$ & flux ($\%$) & $E_{\gamma}$ & $\Delta E_{\gamma}$ & flux ($\%$) \\
      \hline
      \hline
      1 &       Au      &       24.115  &       0.573   &       24.619  &       26.232  &       0.655   &       25.179  &       31.053  &       0.843   &       26.180  &       \\
      & TiO$_{2}$       &       24.111  &       0.596   &       23.525  &       26.227  &       0.688   &       24.089  &       31.040  &       0.933   &       24.850  &       \\
      & Zn      &       24.110  &       0.602   &       23.389  &       26.225  &       0.697   &       23.958  &       31.037  &       0.948   &       24.710  &       \\
      & Os      &       24.109  &       0.609   &       23.159  &       26.222  &       0.714   &       23.620  &       31.034  &       0.968   &       24.491  &       \\
      & Au      &       24.107  &       0.619   &       22.741  &       26.219  &       0.732   &       23.168  &       31.007  &       1.125   &       23.068  &       \\
      \cline{1-11}
      2 &       Au      &       22.987  &       0.830   &       23.308  &       24.912  &       0.948   &       23.239  &       29.245  &       1.264   &       23.080  &       \\
      & TiO$_{2}$       &       22.982  &       0.844   &       21.974  &       24.906  &       0.969   &       22.019  &       29.237  &       1.295   &       22.083  &       \\
      & Zn      &       22.982  &       0.845   &       21.870  &       24.906  &       0.970   &       21.909  &       29.237  &       1.297   &       21.994  &       \\
      & Os      &       22.981  &       0.846   &       21.691  &       24.905  &       0.973   &       21.665  &       29.234  &       1.305   &       21.763  &       \\
      & Au      &       22.980  &       0.847   &       21.369  &       24.904  &       0.977   &       21.316  &       29.220  &       1.362   &       20.687  &       \\
      \cline{1-11}
      3 &       Au      &       21.718  &       0.986   &       18.757  &       23.431  &       1.131   &       18.404  &       27.239  &       1.509   &       17.814  &       \\
      & TiO$_{2}$       &       21.714  &       0.993   &       17.694  &       23.425  &       1.144   &       17.238  &       27.230  &       1.532   &       16.818  &       \\
      & Zn      &       21.714  &       0.993   &       17.599  &       23.425  &       1.144   &       17.150  &       27.229  &       1.533   &       16.726  &       \\
      & Os      &       21.714  &       0.994   &       17.433  &       23.424  &       1.145   &       16.980  &       27.228  &       1.535   &       16.530  &       \\
        & Au      &       21.712  &       0.997   &       16.857  &       23.421  &       1.151   &       16.347  &       27.217  &       1.563   &       15.582  &       \\
    \end{tabular}
  \end{ruledtabular}
\end{table*}




%


\clearpage
\bibliographystyle{apsrev4-2}
\bibliography{citation}

\providecommand{\noopsort}[1]{}\providecommand{\singleletter}[1]{#1}
\begin{thebibliography}{41}%
\makeatletter
\providecommand \@ifxundefined [1]{%
 \@ifx{#1\undefined}
}%
\providecommand \@ifnum [1]{%
 \ifnum #1\expandafter \@firstoftwo
 \else \expandafter \@secondoftwo
 \fi
}%
\providecommand \@ifx [1]{%
 \ifx #1\expandafter \@firstoftwo
 \else \expandafter \@secondoftwo
 \fi
}%
\providecommand \natexlab [1]{#1}%
\providecommand \enquote  [1]{``#1''}%
\providecommand \bibnamefont  [1]{#1}%
\providecommand \bibfnamefont [1]{#1}%
\providecommand \citenamefont [1]{#1}%
\providecommand \href@noop [0]{\@secondoftwo}%
\providecommand \href [0]{\begingroup \@sanitize@url \@href}%
\providecommand \@href[1]{\@@startlink{#1}\@@href}%
\providecommand \@@href[1]{\endgroup#1\@@endlink}%
\providecommand \@sanitize@url [0]{\catcode `\\12\catcode `\$12\catcode
  `\&12\catcode `\#12\catcode `\^12\catcode `\_12\catcode `\%12\relax}%
\providecommand \@@startlink[1]{}%
\providecommand \@@endlink[0]{}%
\providecommand \url  [0]{\begingroup\@sanitize@url \@url }%
\providecommand \@url [1]{\endgroup\@href {#1}{\urlprefix }}%
\providecommand \urlprefix  [0]{URL }%
\providecommand \Eprint [0]{\href }%
\providecommand \doibase [0]{https://doi.org/}%
\providecommand \selectlanguage [0]{\@gobble}%
\providecommand \bibinfo  [0]{\@secondoftwo}%
\providecommand \bibfield  [0]{\@secondoftwo}%
\providecommand \translation [1]{[#1]}%
\providecommand \BibitemOpen [0]{}%
\providecommand \bibitemStop [0]{}%
\providecommand \bibitemNoStop [0]{.\EOS\space}%
\providecommand \EOS [0]{\spacefactor3000\relax}%
\providecommand \BibitemShut  [1]{\csname bibitem#1\endcsname}%
\let\auto@bib@innerbib\@empty
\bibitem [{\citenamefont {DOE}(2015)}]{int1}%
  \BibitemOpen
  \bibfield  {author} {\bibinfo {author} {\bibnamefont {DOE}},\ }\href
  {https://www.osti.gov/biblio/1298983} {\bibfield  {journal} {\bibinfo
  {journal} {2015 NSAC Report}\ } (\bibinfo {year} {2015})}\BibitemShut
  {NoStop}%
\bibitem [{\citenamefont {DOE}(2008)}]{int2}%
  \BibitemOpen
  \bibfield  {author} {\bibinfo {author} {\bibnamefont {DOE}},\ }\href
  {https://science.osti.gov/-/media/Isotope-Research-Development-and-Production/pdf/program/docs/isotope-worksession-presentations/Workshop-Report_final.pdf}
  {\bibfield  {journal} {\bibinfo  {journal} {2008 NSAC Report}\ } (\bibinfo
  {year} {2008})}\BibitemShut {NoStop}%
\bibitem [{\citenamefont {DOE}(2009)}]{int3}%
  \BibitemOpen
  \bibfield  {author} {\bibinfo {author} {\bibnamefont {DOE}},\ }\href
  {https://science.osti.gov/-/media/np/nsac/pdf/docs/nsaci_ii_report.pdf}
  {\bibfield  {journal} {\bibinfo  {journal} {2009 NSAC Report}\ } (\bibinfo
  {year} {2009})}\BibitemShut {NoStop}%
\bibitem [{\citenamefont {Berman}\ and\ \citenamefont {Fultz}(1975)}]{int4}%
  \BibitemOpen
  \bibfield  {author} {\bibinfo {author} {\bibfnamefont {B.~L.}\ \bibnamefont
  {Berman}}\ and\ \bibinfo {author} {\bibfnamefont {S.~C.}\ \bibnamefont
  {Fultz}},\ }\href {https://doi.org/10.1103/RevModPhys.47.713} {\bibfield
  {journal} {\bibinfo  {journal} {Rev. Mod. Phys.}\ }\textbf {\bibinfo {volume}
  {47}},\ \bibinfo {pages} {713} (\bibinfo {year} {1975})}\BibitemShut
  {NoStop}%
\bibitem [{\citenamefont {Berman}(1975)}]{int5}%
  \BibitemOpen
  \bibfield  {author} {\bibinfo {author} {\bibfnamefont {B.}~\bibnamefont
  {Berman}},\ }\href
  {https://doi.org/https://doi.org/10.1016/0092-640X(75)90010-8} {\bibfield
  {journal} {\bibinfo  {journal} {Atomic Data and Nuclear Data Tables}\
  }\textbf {\bibinfo {volume} {15}},\ \bibinfo {pages} {319} (\bibinfo {year}
  {1975})}\BibitemShut {NoStop}%
\bibitem [{\citenamefont {Dietrich}\ and\ \citenamefont {Berman}(1988)}]{int6}%
  \BibitemOpen
  \bibfield  {author} {\bibinfo {author} {\bibfnamefont {S.~S.}\ \bibnamefont
  {Dietrich}}\ and\ \bibinfo {author} {\bibfnamefont {B.~L.}\ \bibnamefont
  {Berman}},\ }\href
  {https://doi.org/https://doi.org/10.1016/0092-640X(88)90033-2} {\bibfield
  {journal} {\bibinfo  {journal} {Atomic Data and Nuclear Data Tables}\
  }\textbf {\bibinfo {volume} {38}},\ \bibinfo {pages} {199} (\bibinfo {year}
  {1988})}\BibitemShut {NoStop}%
\bibitem [{\citenamefont {Koning}\ and\ \citenamefont {the
  others}(2019)\citenamefont {Koning} \emph {et~al.}}]{int7}%
  \BibitemOpen
  \bibfield  {author} {\bibinfo {author} {\bibfnamefont {A.}~\bibnamefont
  {Koning}} \emph {et~al.},\ }\href
  {https://doi.org/https://doi.org/10.1016/j.nds.2019.01.002} {\bibfield
  {journal} {\bibinfo  {journal} {Nuclear Data Sheets}\ }\textbf {\bibinfo
  {volume} {155}},\ \bibinfo {pages} {1} (\bibinfo {year} {2019})},\ \bibinfo
  {note} {special Issue on Nuclear Reaction Data}\BibitemShut {NoStop}%
\bibitem [{\citenamefont {Shibata}\ and\ \citenamefont {the
  others}(2011{\natexlab{a}})\citenamefont {Shibata} \emph {et~al.}}]{int8}%
  \BibitemOpen
  \bibfield  {author} {\bibinfo {author} {\bibfnamefont {K.}~\bibnamefont
  {Shibata}} \emph {et~al.},\ }\href
  {https://doi.org/10.1080/18811248.2011.9711675} {\bibfield  {journal}
  {\bibinfo  {journal} {Journal of Nuclear Science and Technology}\ }\textbf
  {\bibinfo {volume} {48}},\ \bibinfo {pages} {1} (\bibinfo {year}
  {2011}{\natexlab{a}})}\BibitemShut {NoStop}%
\bibitem [{\citenamefont {Shibata}\ and\ \citenamefont {the
  others}(2011{\natexlab{b}})\citenamefont {Shibata} \emph {et~al.}}]{int9}%
  \BibitemOpen
  \bibfield  {author} {\bibinfo {author} {\bibfnamefont {K.}~\bibnamefont
  {Shibata}} \emph {et~al.},\ }\href {https://doi.org/10.3938/jkps.59.1046}
  {\bibfield  {journal} {\bibinfo  {journal} {Journal of the Korean Physical
  Society}\ }\textbf {\bibinfo {volume} {59}},\ \bibinfo {pages} {1046}
  (\bibinfo {year} {2011}{\natexlab{b}})}\BibitemShut {NoStop}%
\bibitem [{\citenamefont {Iwamoto}\ and\ \citenamefont {the
  others}(2011)\citenamefont {Iwamoto} \emph {et~al.}}]{int10}%
  \BibitemOpen
  \bibfield  {author} {\bibinfo {author} {\bibfnamefont {O.}~\bibnamefont
  {Iwamoto}} \emph {et~al.},\ }\href {https://doi.org/10.3938/jkps.59.1224}
  {\bibfield  {journal} {\bibinfo  {journal} {Journal of the Korean Physical
  Society}\ }\textbf {\bibinfo {volume} {59}},\ \bibinfo {pages} {1224}
  (\bibinfo {year} {2011})}\BibitemShut {NoStop}%
\bibitem [{\citenamefont {Chiba}\ and\ \citenamefont {the
  others}(2011)\citenamefont {Chiba} \emph {et~al.}}]{int11}%
  \BibitemOpen
  \bibfield  {author} {\bibinfo {author} {\bibfnamefont {G.}~\bibnamefont
  {Chiba}} \emph {et~al.},\ }\href
  {https://doi.org/10.1080/18811248.2011.9711692} {\bibfield  {journal}
  {\bibinfo  {journal} {Journal of Nuclear Science and Technology}\ }\textbf
  {\bibinfo {volume} {48}},\ \bibinfo {pages} {172} (\bibinfo {year}
  {2011})}\BibitemShut {NoStop}%
\bibitem [{\citenamefont {Chadwick}\ and\ \citenamefont {the
  others}(2011)\citenamefont {Chadwick} \emph {et~al.}}]{int12}%
  \BibitemOpen
  \bibfield  {author} {\bibinfo {author} {\bibfnamefont {M.}~\bibnamefont
  {Chadwick}} \emph {et~al.},\ }\href
  {https://doi.org/https://doi.org/10.1016/j.nds.2011.11.002} {\bibfield
  {journal} {\bibinfo  {journal} {Nuclear Data Sheets}\ }\textbf {\bibinfo
  {volume} {112}},\ \bibinfo {pages} {2887} (\bibinfo {year} {2011})},\
  \bibinfo {note} {special Issue on ENDF/B-VII.1 Library}\BibitemShut {NoStop}%
\bibitem [{\citenamefont {Chadwick}\ and\ \citenamefont {the
  others}(2006)\citenamefont {Chadwick} \emph {et~al.}}]{int13}%
  \BibitemOpen
  \bibfield  {author} {\bibinfo {author} {\bibfnamefont {M.}~\bibnamefont
  {Chadwick}} \emph {et~al.},\ }\href
  {https://doi.org/https://doi.org/10.1016/j.nds.2006.11.001} {\bibfield
  {journal} {\bibinfo  {journal} {Nuclear Data Sheets}\ }\textbf {\bibinfo
  {volume} {107}},\ \bibinfo {pages} {2931} (\bibinfo {year} {2006})},\
  \bibinfo {note} {evaluated Nuclear Data File ENDF/B-VII.0}\BibitemShut
  {NoStop}%
\bibitem [{\citenamefont {Santamirina}\ \emph {et~al.}(2009)\citenamefont
  {Santamirina} \emph {et~al.}}]{int14}%
  \BibitemOpen
  \bibfield  {author} {\bibinfo {author} {\bibfnamefont {A.}~\bibnamefont
  {Santamirina}} \emph {et~al.},\ }\href
  {https://www.oecd-nea.org/jcms/pl_14470} {\bibfield  {journal} {\bibinfo
  {journal} {JEFF Report 22}\ } (\bibinfo {year} {2009})}\BibitemShut {NoStop}%
\bibitem [{\citenamefont {Ge}\ and\ \citenamefont {the
  others}(2011)\citenamefont {Ge} \emph {et~al.}}]{int15}%
  \BibitemOpen
  \bibfield  {author} {\bibinfo {author} {\bibfnamefont {Z.~G.}\ \bibnamefont
  {Ge}} \emph {et~al.},\ }\href {https://doi.org/10.3938/jkps.59.1052}
  {\bibfield  {journal} {\bibinfo  {journal} {Journal of the Korean Physical
  Society}\ }\textbf {\bibinfo {volume} {59}},\ \bibinfo {pages} {1052}
  (\bibinfo {year} {2011})}\BibitemShut {NoStop}%
\bibitem [{\citenamefont {Blokhin}\ and\ \citenamefont {the
  others}(1994)\citenamefont {Blokhin} \emph {et~al.}}]{int16}%
  \BibitemOpen
  \bibfield  {author} {\bibinfo {author} {\bibfnamefont {A.~I.}\ \bibnamefont
  {Blokhin}} \emph {et~al.},\ }\href {https://www.osti.gov/biblio/62095}
  {\bibfield  {journal} {\bibinfo  {journal} {International conference on
  nuclear data for science and technology}\ }\textbf {\bibinfo {volume} {2}}
  (\bibinfo {year} {1994})}\BibitemShut {NoStop}%
\bibitem [{\citenamefont {Titarenko}\ and\ \citenamefont {the
  others}(2008)\citenamefont {Titarenko} \emph {et~al.}}]{int17}%
  \BibitemOpen
  \bibfield  {author} {\bibinfo {author} {\bibfnamefont {Y.~E.}\ \bibnamefont
  {Titarenko}} \emph {et~al.},\ }\href
  {https://doi.org/10.1103/PhysRevC.78.034615} {\bibfield  {journal} {\bibinfo
  {journal} {Phys. Rev. C}\ }\textbf {\bibinfo {volume} {78}},\ \bibinfo
  {pages} {034615} (\bibinfo {year} {2008})}\BibitemShut {NoStop}%
\bibitem [{\citenamefont {Sakane}\ and\ \citenamefont {the
  others}(2002)\citenamefont {Sakane} \emph {et~al.}}]{int18}%
  \BibitemOpen
  \bibfield  {author} {\bibinfo {author} {\bibfnamefont {H.}~\bibnamefont
  {Sakane}} \emph {et~al.},\ }\href
  {https://doi.org/https://doi.org/10.1016/S0306-4549(01)00025-1} {\bibfield
  {journal} {\bibinfo  {journal} {Annals of Nuclear Energy}\ }\textbf {\bibinfo
  {volume} {29}},\ \bibinfo {pages} {53} (\bibinfo {year} {2002})}\BibitemShut
  {NoStop}%
\bibitem [{\citenamefont {Leya}\ and\ \citenamefont {the
  others}(2005)\citenamefont {Leya} \emph {et~al.}}]{int19}%
  \BibitemOpen
  \bibfield  {author} {\bibinfo {author} {\bibfnamefont {I.}~\bibnamefont
  {Leya}} \emph {et~al.},\ }\href
  {https://doi.org/https://doi.org/10.1016/j.nimb.2004.11.009} {\bibfield
  {journal} {\bibinfo  {journal} {Nuclear Instruments and Methods in Physics
  Research Section B: Beam Interactions with Materials and Atoms}\ }\textbf
  {\bibinfo {volume} {229}},\ \bibinfo {pages} {1} (\bibinfo {year}
  {2005})}\BibitemShut {NoStop}%
\bibitem [{\citenamefont {Ma}\ and\ \citenamefont {the
  others}(2015)\citenamefont {Ma} \emph {et~al.}}]{int20}%
  \BibitemOpen
  \bibfield  {author} {\bibinfo {author} {\bibfnamefont {C.-W.}\ \bibnamefont
  {Ma}} \emph {et~al.},\ }\href {https://doi.org/10.1088/0256-307x/32/7/072501}
  {\bibfield  {journal} {\bibinfo  {journal} {Chinese Physics Letters}\
  }\textbf {\bibinfo {volume} {32}},\ \bibinfo {pages} {072501} (\bibinfo
  {year} {2015})}\BibitemShut {NoStop}%
\bibitem [{\citenamefont {Vlastou}\ and\ \citenamefont {the
  others}(2015)\citenamefont {Vlastou} \emph {et~al.}}]{int25}%
  \BibitemOpen
  \bibfield  {author} {\bibinfo {author} {\bibfnamefont {R.}~\bibnamefont
  {Vlastou}} \emph {et~al.},\ }\href
  {https://doi.org/https://doi.org/10.1016/j.phpro.2015.05.053} {\bibfield
  {journal} {\bibinfo  {journal} {Physics Procedia}\ }\textbf {\bibinfo
  {volume} {66}},\ \bibinfo {pages} {425} (\bibinfo {year} {2015})}\BibitemShut
  {NoStop}%
\bibitem [{\citenamefont {Bholane}\ and\ \citenamefont {the
  others}(2021)\citenamefont {Bholane} \emph {et~al.}}]{int26}%
  \BibitemOpen
  \bibfield  {author} {\bibinfo {author} {\bibfnamefont {G.}~\bibnamefont
  {Bholane}} \emph {et~al.},\ }\href
  {https://doi.org/https://doi.org/10.1016/j.apradiso.2021.109739} {\bibfield
  {journal} {\bibinfo  {journal} {Applied Radiation and Isotopes}\ }\textbf
  {\bibinfo {volume} {174}},\ \bibinfo {pages} {109739} (\bibinfo {year}
  {2021})}\BibitemShut {NoStop}%
\bibitem [{\citenamefont {Thiep}\ and\ \citenamefont {the
  others}(2012)\citenamefont {Thiep} \emph {et~al.}}]{int27}%
  \BibitemOpen
  \bibfield  {author} {\bibinfo {author} {\bibfnamefont {T.~D.}\ \bibnamefont
  {Thiep}} \emph {et~al.},\ }\href
  {https://link.springer.com/article/10.1134/S1547477112080092} {\bibfield
  {journal} {\bibinfo  {journal} {Physics of Particles and Nuclei Letters}\
  }\textbf {\bibinfo {volume} {9}},\ \bibinfo {pages} {648} (\bibinfo {year}
  {2012})}\BibitemShut {NoStop}%
\bibitem [{\citenamefont {Plaisir}\ \emph {et~al.}(2012)\citenamefont {Plaisir}
  \emph {et~al.}}]{1}%
  \BibitemOpen
  \bibfield  {author} {\bibinfo {author} {\bibfnamefont {C.}~\bibnamefont
  {Plaisir}} \emph {et~al.},\ }\href
  {https://doi.org/10.1140/epja/i2012-12068-7} {\bibfield  {journal} {\bibinfo
  {journal} {European Physical Journal A: Hadrons and Nuclei}\ }\textbf
  {\bibinfo {volume} {48}},\ \bibinfo {pages} {68} (\bibinfo {year}
  {2012})}\BibitemShut {NoStop}%
\bibitem [{\citenamefont {Fultz}\ \emph {et~al.}(1962)\citenamefont {Fultz}
  \emph {et~al.}}]{2}%
  \BibitemOpen
  \bibfield  {author} {\bibinfo {author} {\bibfnamefont {S.~C.}\ \bibnamefont
  {Fultz}} \emph {et~al.},\ }\href {https://doi.org/10.1103/PhysRev.127.1273}
  {\bibfield  {journal} {\bibinfo  {journal} {Physical Review}\ }\textbf
  {\bibinfo {volume} {127}},\ \bibinfo {pages} {1273} (\bibinfo {year}
  {1962})}\BibitemShut {NoStop}%
\bibitem [{\citenamefont {Sun}\ and\ \citenamefont {Wu}(2011)}]{beam1}%
  \BibitemOpen
  \bibfield  {author} {\bibinfo {author} {\bibfnamefont {C.}~\bibnamefont
  {Sun}}\ and\ \bibinfo {author} {\bibfnamefont {Y.~K.}\ \bibnamefont {Wu}},\
  }\href {https://doi.org/10.1103/PhysRevSTAB.14.044701} {\bibfield  {journal}
  {\bibinfo  {journal} {Phys. Rev. ST Accel. Beams}\ }\textbf {\bibinfo
  {volume} {14}},\ \bibinfo {pages} {044701} (\bibinfo {year}
  {2011})}\BibitemShut {NoStop}%
\bibitem [{\citenamefont {Weller}\ and\ \citenamefont {the
  others}(2009)\citenamefont {Weller} \emph {et~al.}}]{int23}%
  \BibitemOpen
  \bibfield  {author} {\bibinfo {author} {\bibfnamefont {H.~R.}\ \bibnamefont
  {Weller}} \emph {et~al.},\ }\href
  {https://doi.org/https://doi.org/10.1016/j.ppnp.2008.07.001} {\bibfield
  {journal} {\bibinfo  {journal} {Progress in Particle and Nuclear Physics}\
  }\textbf {\bibinfo {volume} {62}},\ \bibinfo {pages} {257} (\bibinfo {year}
  {2009})}\BibitemShut {NoStop}%
\bibitem [{\citenamefont {Vogt}\ \emph {et~al.}(2002)\citenamefont {Vogt} \emph
  {et~al.}}]{9}%
  \BibitemOpen
  \bibfield  {author} {\bibinfo {author} {\bibfnamefont {K.}~\bibnamefont
  {Vogt}} \emph {et~al.},\ }\href
  {https://doi.org/10.1016/S0375-9474(02)00922-3} {\bibfield  {journal}
  {\bibinfo  {journal} {Nuclear Physics, Section A}\ }\textbf {\bibinfo
  {volume} {707}},\ \bibinfo {pages} {241} (\bibinfo {year}
  {2002})}\BibitemShut {NoStop}%
\bibitem [{\citenamefont {Berman}\ \emph {et~al.}(1987)\citenamefont {Berman}
  \emph {et~al.}}]{8}%
  \BibitemOpen
  \bibfield  {author} {\bibinfo {author} {\bibfnamefont {B.~L.}\ \bibnamefont
  {Berman}} \emph {et~al.},\ }\href {https://doi.org/10.1103/PhysRevC.36.1286}
  {\bibfield  {journal} {\bibinfo  {journal} {Physical Review, Part C, Nuclear
  Physics}\ }\textbf {\bibinfo {volume} {36}},\ \bibinfo {pages} {1286}
  (\bibinfo {year} {1987})}\BibitemShut {NoStop}%
\bibitem [{\citenamefont {Veyssiere}\ \emph {et~al.}(1970)\citenamefont
  {Veyssiere} \emph {et~al.}}]{7}%
  \BibitemOpen
  \bibfield  {author} {\bibinfo {author} {\bibfnamefont {A.}~\bibnamefont
  {Veyssiere}} \emph {et~al.},\ }\href
  {https://doi.org/10.1016/0375-9474(70)90727-X} {\bibfield  {journal}
  {\bibinfo  {journal} {Nuclear Physics, Section A}\ }\textbf {\bibinfo
  {volume} {159}},\ \bibinfo {pages} {561} (\bibinfo {year}
  {1970})}\BibitemShut {NoStop}%
\bibitem [{\citenamefont {Itoh}\ \emph {et~al.}(2011)\citenamefont {Itoh} \emph
  {et~al.}}]{5}%
  \BibitemOpen
  \bibfield  {author} {\bibinfo {author} {\bibfnamefont {O.}~\bibnamefont
  {Itoh}} \emph {et~al.},\ }\href
  {https://doi.org/10.1080/18811248.2011.9711766} {\bibfield  {journal}
  {\bibinfo  {journal} {Jour. of Nuclear Science and Technology}\ }\textbf
  {\bibinfo {volume} {48}},\ \bibinfo {pages} {834} (\bibinfo {year}
  {2011})}\BibitemShut {NoStop}%
\bibitem [{\citenamefont {Kitatani}\ \emph {et~al.}(2011)\citenamefont
  {Kitatani} \emph {et~al.}}]{6}%
  \BibitemOpen
  \bibfield  {author} {\bibinfo {author} {\bibfnamefont {F.}~\bibnamefont
  {Kitatani}} \emph {et~al.},\ }\href {https://doi.org/10.3327/jnst.48.1017}
  {\bibfield  {journal} {\bibinfo  {journal} {Jour. of Nuclear Science and
  Technology}\ }\textbf {\bibinfo {volume} {48}},\ \bibinfo {pages} {1017}
  (\bibinfo {year} {2011})}\BibitemShut {NoStop}%
\bibitem [{\citenamefont {Kitatani}\ \emph {et~al.}(2010)\citenamefont
  {Kitatani} \emph {et~al.}}]{4}%
  \BibitemOpen
  \bibfield  {author} {\bibinfo {author} {\bibfnamefont {F.}~\bibnamefont
  {Kitatani}} \emph {et~al.},\ }\href {https://doi.org/10.3327/jnst.47.367}
  {\bibfield  {journal} {\bibinfo  {journal} {Jour. of Nuclear Science and
  Technology}\ }\textbf {\bibinfo {volume} {47}},\ \bibinfo {pages} {367}
  (\bibinfo {year} {2010})}\BibitemShut {NoStop}%
\bibitem [{\citenamefont {Hara}\ \emph {et~al.}(2007)\citenamefont {Hara} \emph
  {et~al.}}]{3}%
  \BibitemOpen
  \bibfield  {author} {\bibinfo {author} {\bibfnamefont {K.~Y.}\ \bibnamefont
  {Hara}} \emph {et~al.},\ }\href {https://doi.org/10.3327/jnst.44.938}
  {\bibfield  {journal} {\bibinfo  {journal} {Jour. of Nuclear Science and
  Technology}\ }\textbf {\bibinfo {volume} {44}},\ \bibinfo {pages} {938}
  (\bibinfo {year} {2007})}\BibitemShut {NoStop}%
\bibitem [{\citenamefont {Yokoya}(2003)}]{beam2}%
  \BibitemOpen
  \bibfield  {author} {\bibinfo {author} {\bibfnamefont {K.}~\bibnamefont
  {Yokoya}},\ }\href@noop {} {\bibfield  {journal} {\bibinfo  {journal} {User
  manual of $\textsc{CAIN}$, version 2.35}\ } (\bibinfo {year}
  {2003})}\BibitemShut {NoStop}%
\bibitem [{\citenamefont {Hubbell}(1982)}]{bf1}%
  \BibitemOpen
  \bibfield  {author} {\bibinfo {author} {\bibfnamefont {J.}~\bibnamefont
  {Hubbell}},\ }\href
  {https://doi.org/https://doi.org/10.1016/0020-708X(82)90248-4} {\bibfield
  {journal} {\bibinfo  {journal} {The International Journal of Applied
  Radiation and Isotopes}\ }\textbf {\bibinfo {volume} {33}},\ \bibinfo {pages}
  {1269} (\bibinfo {year} {1982})}\BibitemShut {NoStop}%
\bibitem [{\citenamefont {Hubbell}\ and\ \citenamefont {Seltzer}(1995)}]{bf2}%
  \BibitemOpen
  \bibfield  {author} {\bibinfo {author} {\bibfnamefont {J.}~\bibnamefont
  {Hubbell}}\ and\ \bibinfo {author} {\bibfnamefont {S.}~\bibnamefont
  {Seltzer}},\ }\href
  {http://physics.nist.gov/PhysRefData/XrayMassCoef/cover.html} {\bibfield
  {journal} {\bibinfo  {journal} {X-Ray Mass Attenuation Coefficients}\ }
  (\bibinfo {year} {1995})}\BibitemShut {NoStop}%
\bibitem [{\citenamefont {Pywell}\ \emph {et~al.}(2009)\citenamefont {Pywell}
  \emph {et~al.}}]{bf3}%
  \BibitemOpen
  \bibfield  {author} {\bibinfo {author} {\bibfnamefont {R.}~\bibnamefont
  {Pywell}} \emph {et~al.},\ }\href
  {https://doi.org/https://doi.org/10.1016/j.nima.2009.04.014} {\bibfield
  {journal} {\bibinfo  {journal} {Nuclear Instruments and Methods in Physics
  Research Section A: Accelerators, Spectrometers, Detectors and Associated
  Equipment}\ }\textbf {\bibinfo {volume} {606}},\ \bibinfo {pages} {517}
  (\bibinfo {year} {2009})}\BibitemShut {NoStop}%
\bibitem [{\citenamefont {Radford}(1995)}]{exp1}%
  \BibitemOpen
  \bibfield  {author} {\bibinfo {author} {\bibfnamefont {D.}~\bibnamefont
  {Radford}},\ }\href
  {https://doi.org/https://doi.org/10.1016/0168-9002(95)00183-2} {\bibfield
  {journal} {\bibinfo  {journal} {Nucl. Instrum. Methods Phys. Res. Sect. A}\
  }\textbf {\bibinfo {volume} {361}},\ \bibinfo {pages} {297} (\bibinfo {year}
  {1995})}\BibitemShut {NoStop}%
\bibitem [{\citenamefont {National Nuclear Data~Center}(2008)}]{Au1}%
  \BibitemOpen
  \bibfield  {author} {\bibinfo {author} {\bibfnamefont {B.~N.~L.}\
  \bibnamefont {National Nuclear Data~Center}},\ }\href
  {https://www.nndc.bnl.gov/nudat3/} {\bibfield  {journal} {\bibinfo  {journal}
  {www.nndc.bnl.gov}\ } (\bibinfo {year} {2008})}\BibitemShut {NoStop}%
\bibitem [{\citenamefont {Iwase}\ \emph {et~al.}(2002)\citenamefont {Iwase}
  \emph {et~al.}}]{beam3}%
  \BibitemOpen
  \bibfield  {author} {\bibinfo {author} {\bibfnamefont {H.}~\bibnamefont
  {Iwase}} \emph {et~al.},\ }\href
  {https://doi.org/10.1080/18811248.2002.9715305} {\bibfield  {journal}
  {\bibinfo  {journal} {Journal of Nuclear Science and Technology}\ }\textbf
  {\bibinfo {volume} {39}},\ \bibinfo {pages} {1142} (\bibinfo {year}
  {2002})}\BibitemShut {NoStop}%
\end{thebibliography}%

\end{document}